\documentclass[useAMS,usenatbib]{mn2e}

\usepackage{epsfig}
\usepackage{textcomp}
\usepackage{longtable}
\usepackage{lscape}

\newcommand{\xmm} {{\it XMM-Newton}}
\newcommand{\chandra} {{\it Chandra}}
\newcommand{\cmsq} {cm$^{-2}$}
\newcommand{\nh} {$N_{\rm{H}}$}
\newcommand{\lx} {$L_{\rm{X}}$}
\newcommand{\chisq} {$\chi^2$}

\newcommand{\oiii}{{\rm{[O\,\sc{iii}]}}}

\newcommand{\degree}{{$^\circ$}}
\newcommand{\ergs}{\mbox{\thinspace erg\thinspace s$^{-1}$}}
\newcommand{\fscat}{$f_{\rm scatt}$}

\title[]{The evolution of the Compton thick fraction and the nature of obscuration for AGN in the \chandra\ Deep Field South}

\author[M. Brightman and Y. Ueda]{Murray Brightman$^{1,2}\thanks{E-mail: mbright@mpe.mpg.de}$ and Yoshihiro Ueda$^{1}$\\
$^{1}$Department of Astronomy, Kyoto University, Kyoto 606-8502, Japan\\
$^{2}$Max-Planck-Institut f\"{u}r extraterrestrische Physik, Giessenbachstrasse 1, D-85748, Garching bei M\"{u}nchen, Germany\\}

\begin{document}

\date{Accepted 0000 December 00. Received 0000 December 00; in original form 0000 October 00}

\pagerange{\pageref{firstpage}--\pageref{lastpage}} \pubyear{0000}

\maketitle

\label{firstpage}

\begin{abstract}
We present the results from the X-ray spectral analysis of high redshift active galactic nuclei (AGN) in the Chandra Deep Field-South (CDFS), making use of the new 4Ms data set and new X-ray spectral models from Brightman \& Nandra, which account for Compton scattering and the geometry of the circumnuclear material. Our goals are to ascertain to what extent the torus paradigm of local AGN is applicable at earlier epochs and to evaluate the evolution of the Compton thick fraction with redshift, important for X-ray background synthesis models and understanding the accretion history of the universe. In addition to the torus models, we measure the fraction of scattered nuclear light, \fscat, known to be dependant on covering factor of the circumnuclear materal, and use this to aid in our understanding of its geometry. We find that the covering factor of the circumnuclear material is correlated with the line of sight \nh, and as such the most heavily obscured AGN are in fact also the most geometrically buried. We come to these conclusions from the result that \fscat\ decreases as \nh\ increases and from the prevalence of the torus model with the smallest opening angle as best fit model in the fits to the most obscured AGN. We find that a significant fraction of sources ($\sim20$\%) in the CDFS  are likely to be buried in material with close to 4$\pi$ coverage having been best fit by the torus model with a 0\degree\ opening angle. Furthermore, we find 41 Compton thick sources in the CDFS using the new torus models, 29 of which we report here for the first time. We bin our sample by redshift in order to investigate the evolution of the Compton thick fraction by epoch. We take into account the incompleteness and contamination rates in the spectral identification of Compton thick AGN using data from simulations. We also account for the bias against the X-ray selection of heavily obscured sources due to flux suppression in the \chandra\ band, by restricting to intrinsic luminosities to which the CDFS is sensitive to Compton thick AGN (\lx$\sim10^{44}$ at z=2.5). We find a significant increase in the intrinsic Compton thick fraction, normalised to \lx$=10^{43.5}$ \ergs, from $\approx20$\% in the local universe to $\approx$40\% at z=1-4.
\end{abstract}


\section{Introduction}

It is well known that a large fraction of AGN are obscured by gas and dust.  This has been seen both observationally by the measurement of line of sight absorption in X-ray spectra \citep{awaki91,risaliti99}, and has also been inferred, due to a large population of obscured AGN being required to fit the cosmic X-ray background \citep{comastri95,ueda03,gilli07}. The obscuration is, for the most part, attributable to circumnuclear material surrounding the black hole on parsec scales. In the local universe, a unification paradigm exists, which is invoked to explain the differences between type 1 AGN, in which broad emission lines, emitted from gas orbiting close to the black hole, can be seen in the optical spectra, and type 2 AGN, in which the broad lines are not directly visible, but are often visible in scattered emission \citep{antonucci93,urry95,tran03}. This unification scheme describes a torus-like structure, which obscures the central engine for some sight lines, and not for others, producing the two observed AGN types, depending on the orientation of the observer. The torus paradigm of the unified scheme has much observational support in the local universe, but the relevance of the unified scheme at higher redshifts is not so clear \citep{draper11,page11}. Characterising AGN obscuration, especially at intermediate redshifts and the most extremely obscured, Compton thick (\nh$\geq1.5\times10^{24}$\cmsq) AGN population is key in understanding the accretion history of the universe \citep{fabian99}.

Much can be learned about AGN obscuration through X-ray spectroscopy. Not only can X-rays penetrate all but the most severe obscuration, but the signature of obscuration is imprinted on the X-ray spectrum via soft X-ray attenuation and the emission of the iron K$\alpha$ fluorescence line. Work to characterise obscuration for high redshift sources through X-ray spectral analysis has been carried out by \cite{tozzi06} using the 1Ms \chandra\ Deep Field South (CDFS) data and \cite{georgantopoulos09} using the 2Ms \chandra\ Deep Field North (CDFN) data. These two fields provide the deepest pictures of the X-ray universe, with the latest observations in the CDFS bringing the total exposure in the field to 4Ms, thereby reaching on axis sensitivities of $\sim10^{-17}$ \ergs\ \citep{xue11}. Both \cite{tozzi06} and \cite{georgantopoulos09} present sets of Compton thick candidates from their analysis, but recently more fiduciary identification of a hand full of Compton thick sources in the CDFS have been presented by \cite{comastri11} using deep \xmm\ data, as well as by \cite{feruglio11} and \cite{gilli11} using the 4Ms \chandra\ data.

In X-ray spectral analysis it is commonplace to use models which attribute X-ray absorption to photoelectric absorption along the line of sight (e.g. {\it zwabs} in {\sc xspec}, see \cite{mainieri07} for an example of this.).  This approach is sufficient at low column densities, however, when the line of sight column density (\nh) reaches $\sim10^{23}$ \cmsq, the optical depth to Compton scattering becomes non-negligible. Not only must attenuation of the X-rays by Compton scattering be accounted for at this point, but the geometry of the obscuring material also plays an important role in the shape of the observed  X-ray spectrum. For example, in Compton thick sources, the direct transmitted emission may not be visible at all, but instead scattered nuclear emission can be seen, either by Compton reflection from the optically thick wall of the torus, or by Thompson scattering from hot gas that fills the torus. By considering these scattering effects, not only can X-ray spectra indicate the level of absorption occurring, but they can also give clues about the geometry of the circumnuclear material.

Recently, several works have presented new X-ray spectral models which properly account for Compton scattering and the geometry of the obscuring material \citep{murphy09, ikeda09, brightman11}. These models make use of Monte-Carlo methods to simulate the non-linear nature of the scattering, and include self consistent iron K$\alpha$ emission. \cite{murphy09} present a model which considers a torus with a circular cross section, with the opening angle fixed at 60$^{\circ}$, whereas \cite{ikeda09} present a model for a spherical torus (a sphere with a biconical structure removed), with a variable opening angle, but where the line of sight \nh\ varies with the viewing angle. The model we use for this work is that presented by \cite{brightman11}, henceforth BN11, being a spherical torus with a varying opening angle, but a constant line of sight \nh.

Some progress has been made at using X-ray spectral information to constrain the geometry of the obscuring material in AGN. For example, \cite{awaki09}, \cite{eguchi11} and \cite{tazaki11} have applied the model of \cite{ikeda09} to several broad band {\it Suzaku} spectra of absorbed AGN, managing to place constraints on the torus geometrical parameters. Furthermore \cite{rivers11} have used the \cite{murphy09} model to constrain the Compton thickness of the circumnuclear material from an unobscured AGN, also using {\it Suzaku}. It is not only efforts using these new spectral models that have helped determine the geometry of the circum-nuclear material in AGN using X-ray spectral data. \cite{ueda07} announced the discovery of a new type of buried AGN in which the circum-nuclear material is likely to be geometrically thick with a high covering fraction. This conclusion was reached by the measurement of a very small ($<0.5\%$) scattered fraction, \fscat. This so called ``scattered'' emission is thought to originate from the Thompson scattering of the primary X-ray photons by hot electrons within the cone of the torus, and is ubiquitous in obscured AGN \citep{turner97}, or emission from circum-nuclear plasma photo-ionised by the AGN \citep{guainazzi07}. A small scattered fraction implies a small opening angle of the torus, or otherwise an under abundance of the gas responsible for the scattering. This picture is supported by \cite{noguchi10} who found that the scattered fraction correlates with the \oiii\ narrow emission line to X-ray flux ratio. As the \oiii\ emission line also originates in gas responsible for the scattering of the X-rays, it is also gives an indication of the covering factor of the gas, or its abundance.

While these works have concentrated on a small set of high quality X-ray spectra of local AGN, the aim of this paper is to take advantage of the 4Ms CDFS data, which has resulted in many spectra containing of order of hundreds of counts or more, enabling application of more such sophisticated models. Our goal is to use both these new X-ray spectral models and the measurement of \fscat\ to characterise the nature of the obscuring material in AGN in the CDFS, and to use them to identify Compton thick sources. In section \ref{chandata} we describe our data set and the data reduction techniques, in section \ref{specanalysis} we describe our spectral fitting technique using the new torus models of BN11, and in section \ref{results} we present our results on our investigation into the geometry of the absorber in high redshift AGN and identification of Compton thick sources. We discuss these results in section \ref{discussion}. In this work we assume a flat cosmological model with $H_{\rm 0}$=70.5 km s$^{-1}$ Mpc$^{-1}$ and $\Omega_{\Lambda}$=0.73 \citep{komatsu09}. For measurement uncertainties on our spectral fit parameters we present the 90\% confidence limits given two interesting parameters.

\section{Chandra data}
\label{chandata}

The \chandra\ Deep Field South currently consists of 52 individual observations, giving $\sim$4 Ms of total exposure time in the field. Of these observations, nine were made in 2000 \citep[$\sim$1 Ms,][]{giacconi02}, 12 made in 2007 \citep[$\sim$1 Ms,][]{luo08} and a further 31 made in 2010 \citep[$\sim$2 Ms,][]{xue11}. We analyse the entire 4Ms data set, where data were screened for hot pixels and cosmic afterglows as described in \cite{laird09}. For our spectral analysis, we use the source list presented by \cite{luo08}, based on detections in the 2Ms data set, consisting of 462 sources down to a flux limit of $\sim1\times10^{-16}$ ergs cm$^{-2}$ s$^{-1}$ in the 2-8 keV band and $\sim2\times10^{-17}$ ergs cm$^{-2}$ s$^{-1}$ in the 0.5-2 keV band. As the focus of this work in on X-ray spectral analysis, we neglect the faintest sources newly detected in the 4Ms data set presented by \cite{xue11}. Events were extracted from the data using the {\sc acis extract} (AE) software package \footnote{The {\sc acis extract} software package and User's Guide are available at http://www.astro.psu.edu/xray/acis/acis\_analysis.html} \citep{broos10}. AE extracts spectral information for each source from each individual observation ID (obsID) based on the shape of the local point spread function (PSF) for that particular position on the detector. We choose to use regions where 90\% of the PSF has been enclosed at 1.5 keV. Background spectra are extracted from an events list which has been masked of all 776 detected point sources in \cite{xue11}, using regions which contain at least 100 counts. AE also constructs response matrix files (RMF) and auxiliary matrix files (ARF). The data from each obsID is then merged to create a single source spectrum, background spectrum, RMF and ARF for each source. 

\begin{figure}
\includegraphics[width=90mm]{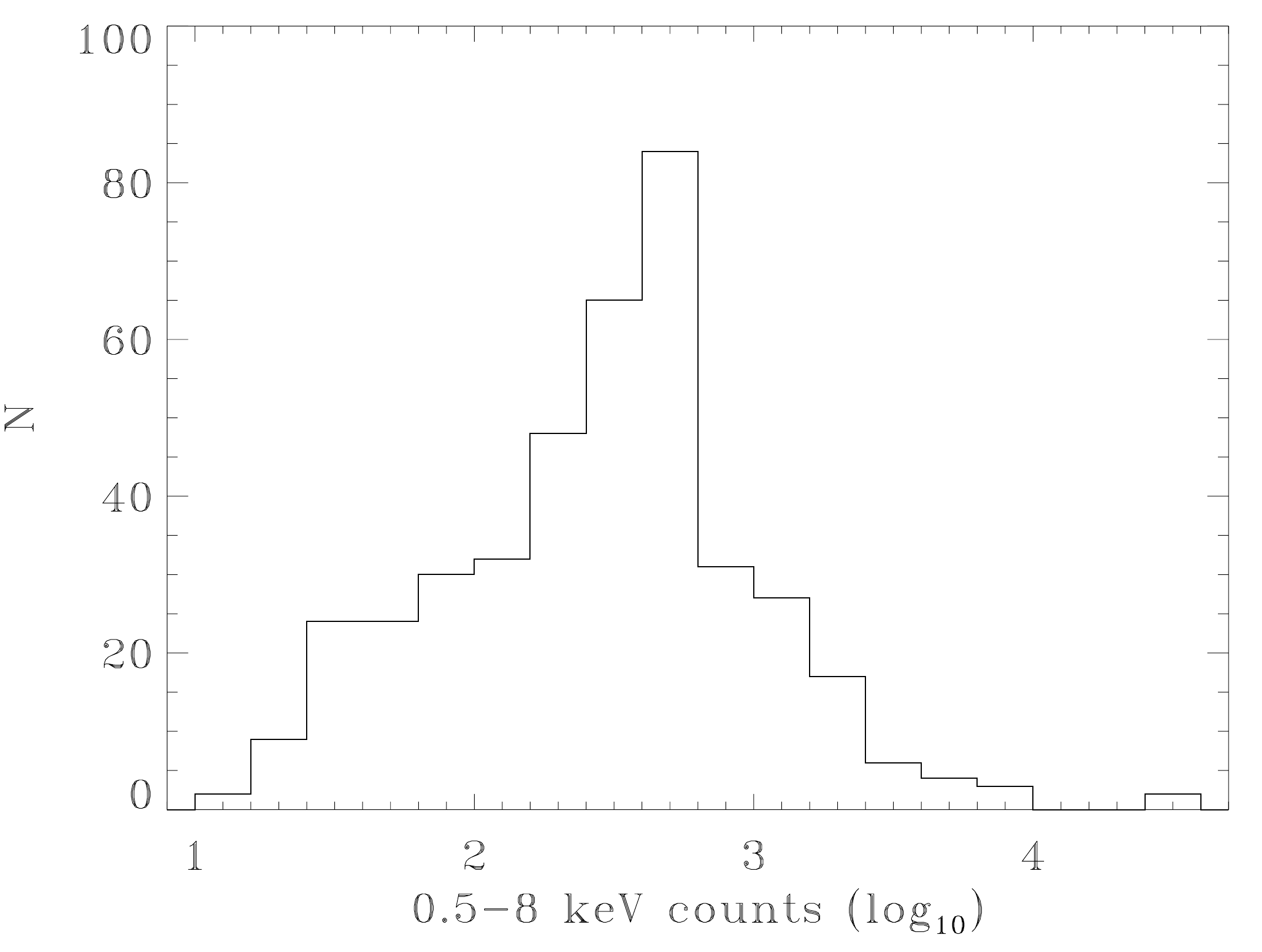}

\caption{A histogram of the total number of 0.5-8 keV spectral data counts in our sample.}
 \label{fig_cnts}
\end{figure}

\begin{figure}
\includegraphics[width=90mm]{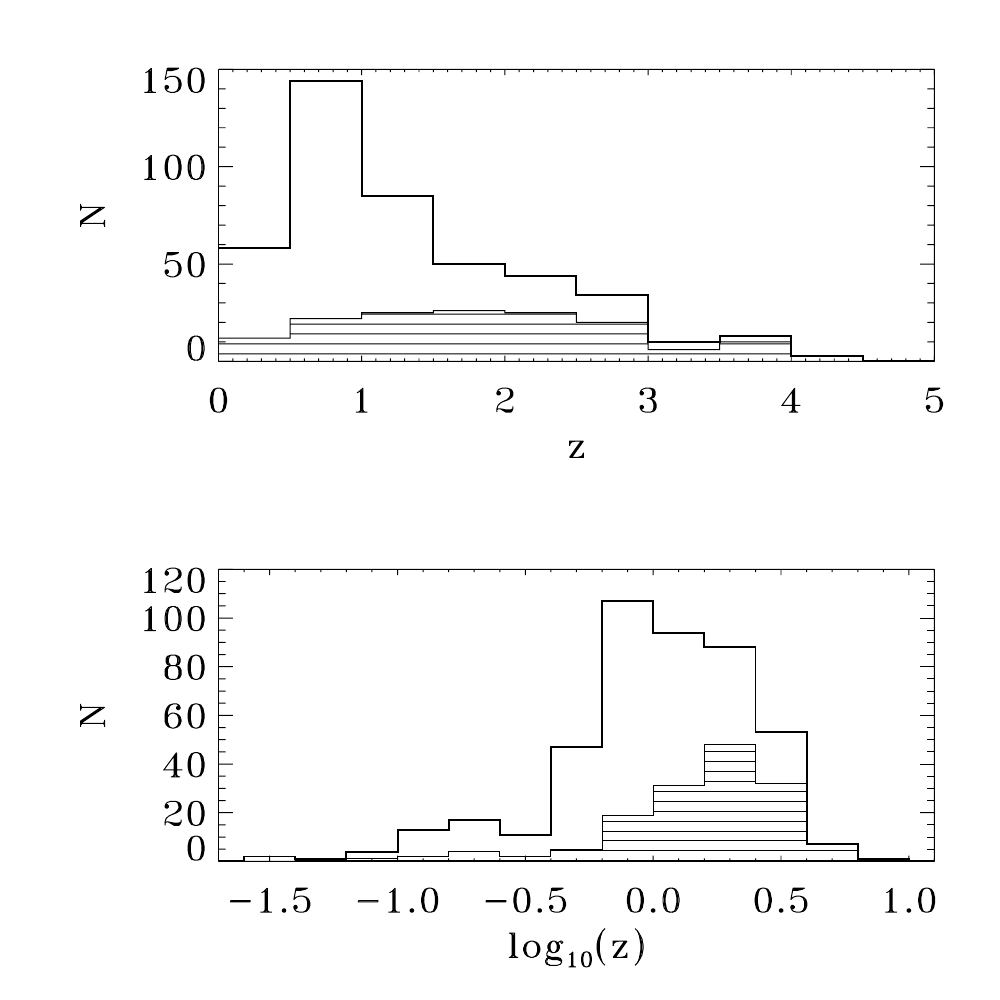}

\caption{Histograms of the redshifts of the sources in our sample, shown with both linear and logarithmic binning. The hatched histograms within the empty histograms show the photometric redshifts used.}
 \label{fig_red}
\end{figure}

Of the 462 sources presented by \cite{luo08}, we use 408 secure spectroscopic or photometric redshifts as compiled by \cite{silverman10} using new data from the VLT and Keck as well as archival data from \cite{szokoly04}, \cite{lefevre04}, \cite{wolf04}, \cite{zheng04}, \cite{ravikumar07}, \cite{wolf08}, \cite{popesso09}, \cite{treister09} and \cite{balestra10}. We use a further 41  photometric redshifts presented in \cite{xue11} for sources not presented in \cite{silverman10}. The total number of sources for which we have obtained a redshift for is 449, of which 296 are spectroscopic redshifts. Considering both spectroscopic and photometric redshifts, the redshift identification completeness of our sample is 97\%. For the remaining 13 sources without redshifts, we neglect them from our analysis. Fig. \ref{fig_cnts} shows the distribution of spectral counts for our sample, and Fig. \ref{fig_red} shows their redshift distribution.

\section{X-ray spectral analysis}
\label{specanalysis}

\subsection{X-ray spectral fitting}

We carefully assess our options for our spectral fitting technique. Due to relatively low count statistics in the CDFS spectra, and our wish to preserve as much spectral resolution as possible, we do not wish to significantly group our spectra. The standard fit statistic used in X-ray spectral analysis is the \chisq\ statistic. However, this is known to fail when the spectrum contains bins in which there are only a few counts, making it necessary to bin the spectra, usually with a minimum of 20 counts per bin. The fit statistic most appropriate for fitting with low numbers of counts per bin is the Cash statistic \citep[cstat,][]{cash79} which uses a Poisson likelihood function. For the spectral fitting in the CDFS carried out by \cite{tozzi06}, the authors carried out simulations of low count spectra to compare results from fitting using the Cash statistic and the \chisq\ statistic. They find that not only does the Cash statistic produce a result closer to the input parameters, but the rms dispersion on the result is also lower when using Cash statistics. They also emphasise the point that as the \chisq\ statistic requires that the spectrum be binned, the spectrum used has essentially been smoothed removing fine spectra details, such as emission lines. The Cash statistic is not normally suitable for background subtracted spectra, nevertheless, {\sc xspec} has a modified version of cstat for when a background spectrum is included\footnote{https://astrophysics.gsfc.nasa.gov/XSPECwiki/low\_count\_spectra}. We use {\sc xspec} version 12.6.0q to carry out X-ray spectral fitting. For these reasons, we have selected to use the Cash statistic in our analysis, where we only lightly group our spectra with at least one count per bin as suggested in the documentation, using the {\sc heasarc} tool {\tt grppha}.

\subsection{Determining the geometry of the absorber}

\begin{figure}
\begin{center}
\includegraphics[width=90mm]{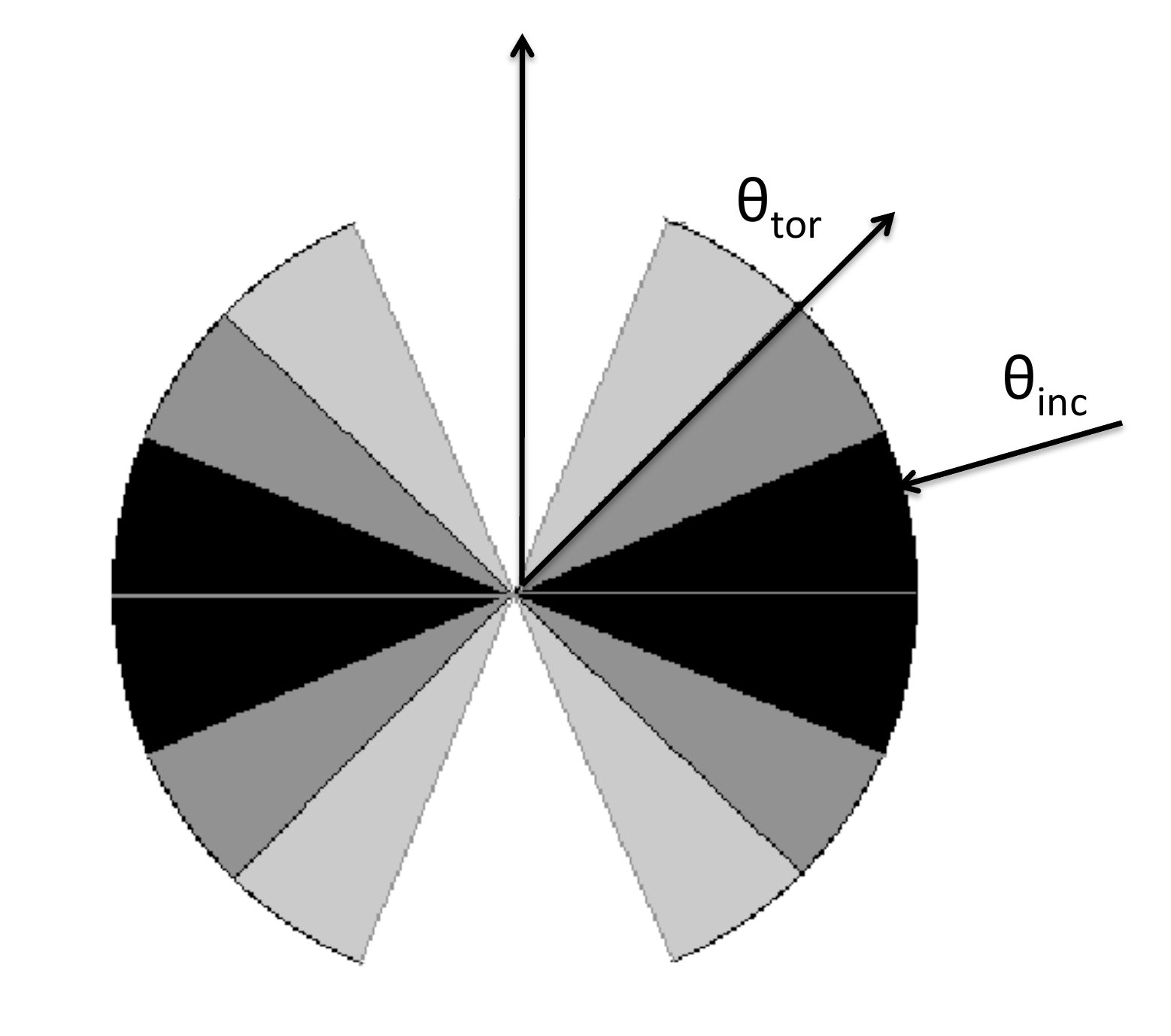}
\includegraphics[width=90mm]{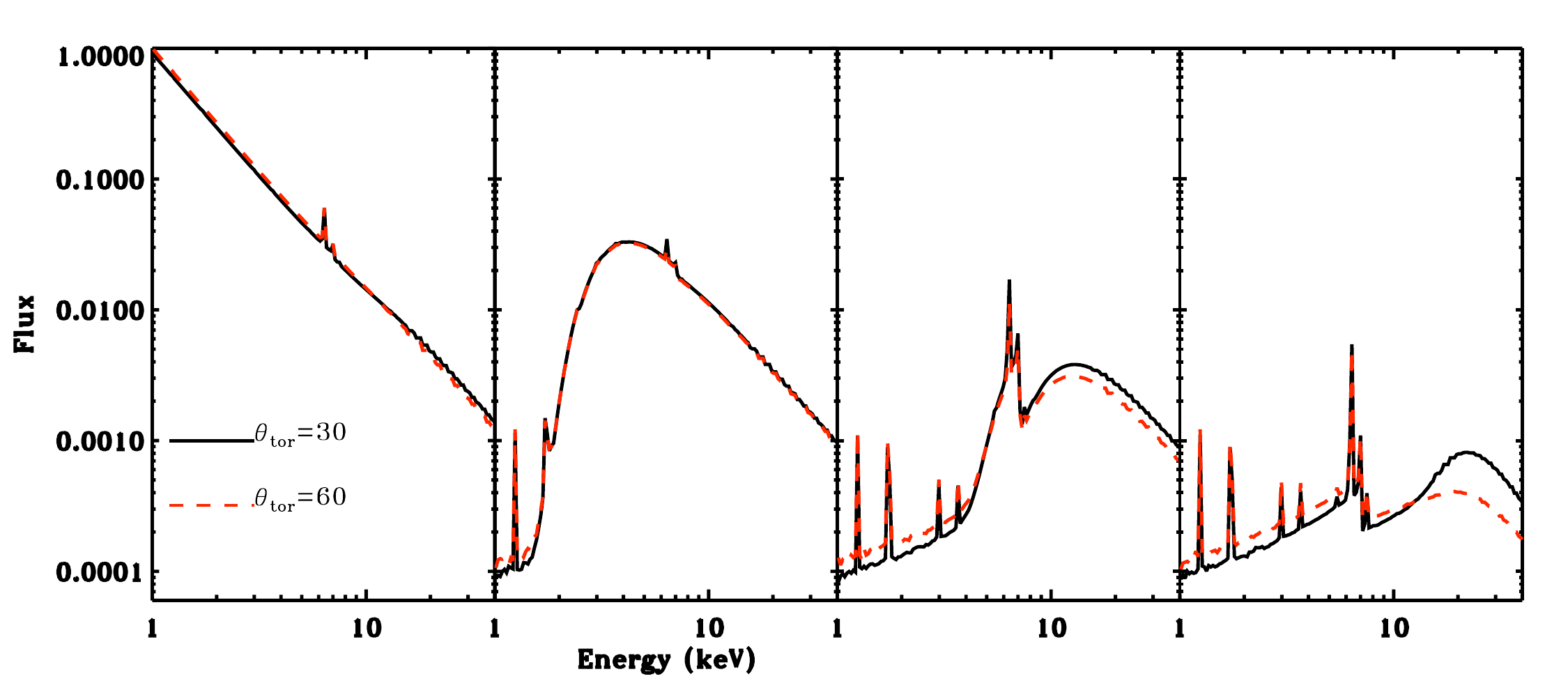}

\caption{Geometry of the torus model presented in BN11. The shading only serves to differentiate the different opening angles used, and does not correspond to density. Typical spectra are shown below the diagram. The first panel shows the unobscured sight-line ($\theta_{inc}$=20\degree), whereas the other panels show obscured sight-lines (\nh=$10^{23}$ \cmsq, $\theta_{inc}$=80\degree; \nh=$10^{24}$ \cmsq, $\theta_{inc}$=80\degree\ and \nh=$10^{25}$ \cmsq, $\theta_{inc}$=80\degree\ respectively). The solid black lines are for $\theta_{tor}$=30\degree\ and the dashed red lines are for $\theta_{tor}$=60\degree}
\label{fig_torgeom}
\end{center}
% \label{fig:xrb}
\end{figure}

A major motivation behind this work is to assess the geometry of the obscuring material in AGN in the CDFS using X-ray spectroscopy. For this we use the new torus models described in BN11. These models are based on Monte-Carlo simulations of an X-ray point source, emitting a power-law spectrum, $F_E=AE^{-\Gamma}$ (range $1.0\leq\Gamma\leq3.0$) at the centre of a distribution of a cold neutral medium, with a line of sight column density to the observer, \nh\ (range $10^{20}\leq$\nh$\leq10^{26}$ \cmsq). The interactions modelled are photoelectric absorption, Compton scattering, and K-shell fluorescence from several elements, including Fe K$\alpha$ and Fe K$\beta$ emission, where the abundances of the elemental constituents are assumed to be solar. The geometry of the obscuring material modelled is a spherical torus which is essentially a sphere, with a biconical cone cut into it.  The geometrical parameters of this distribution are the half opening angle of the torus ($25.8^{\circ} \le \theta_{tor} \le 84.3^{\circ}$) and the inclination angle of the torus ($18.2^{\circ} \le \theta_{i} \le 87.1^{\circ}$). Fig. \ref{fig_torgeom} shows the geometry of the torus, and shows how $\theta_{tor}$ and $\theta_{i}$ are defined. For sight lines through the torus, the \nh\ is independent of the viewing angle.  Also presented in BN11 is a special case, whereby the  source is covered by 4$\pi$ steradians of material in a spherical distribution.

It is unlikely however, that the torus geometrical parameters can be constrained for many individual faint sources in the CDFS. Therefore, in our spectral analysis, our goal is to investigate the prevalence of the best fit model, considering three geometrical scenarios, where the geometrical parameters have been fixed. These are a torus geometry, where the opening angle is fixed to either 60\degree, 30\degree\ or 0\degree\ (i.e spherical) and the inclination angle is fixed to edge on. For determining the geometry of the absorber, we fit only the 408 sources with the most secure redshifts, as presented in \cite{silverman10} so as to minimise the effect from uncertainties in the photometric redshifts. We fit each of these source spectra with the following five combinations of spectral models, which are physically motivated and represent different geometrical scenarios for the circumnuclear material. For the torus scenarios where the opening angle is $>$0\degree, we include a secondary power-law in the fit, which represents a fraction of the nuclear emission posited to be scattered into the line of sight by gas filling the cone of the torus. 

\begin{itemize}

\item{A) The torus model of BN11 with a fixed opening angle of 60$^\circ$, and edge on orientation, accompanied by a scattered power-law, with $\Gamma_{\rm scatt}$ fixed to the value of the primary power-law. This model has four free parameters, \nh; $\Gamma$; and the normalisations of the power-laws, A$_1$ and A$_2$}
\item{B) The torus model of BN11 with a fixed opening angle of 30$^\circ$, and edge on orientation, accompanied by a scattered power-law, with $\Gamma$ fixed to the value of the primary power-law. This model has four free parameters, \nh; $\Gamma$; A$_1$ and A$_2$}
\item{C) The spherical model of BN11, a scenario in which the X-ray source is completely covered with 4$\pi$ steradians of obscuring material. This model has three free parameters,  \nh; $\Gamma$ and A$_1$}. We include no scattered component for this model as it represents the case where there is no escape route for the primary radiation to be scattered into the line of sight. We do however calculate an upper limit on this component.

The new torus models take the place of well established models used previously to model absorption (e.g. {\it zwabs}) and reflection from Compton thick material (e.g. {\it pexrav}). The model {\it plcabs} \citep{yaqoob97} describes the special case where the source is covered by 4$\pi$ steradians of obscuring material, analogous to our spherical model.

The final two models describe unobscured X-ray emission, though these sources are not the focus of our work.
\item{D) simple power-law model, with two free parameters, the power-law index, $\Gamma$ and the normalisation, A$_1$, representing unobscured X-ray emission.}
\item{E) The torus model of BN11 with a fixed opening angle of 60$^\circ$, and face on orientation. This model is essentially a simple power-law, but includes the reflection features from the torus, such as the Fe K$\alpha$ line and Compton hump. This model has three free parameters, \nh\ (through the torus); $\Gamma$ and A$_1$}
\end{itemize}

We choose the model which presents the best c-stat value, after accounting for the number of free parameters of the model. For example, for model C, with three free parameters to be chosen as best fit model over model D, with two free parameters, it must present a c-stat improvement of at least 1.0, where 1.0 is the $\Delta$c-stat associated with the addition of a free parameter at the 1-$\sigma$ confidence level. i.e. cstat(C)$<$cstat(D)-1.0

\begin{figure}
\includegraphics[width=90mm]{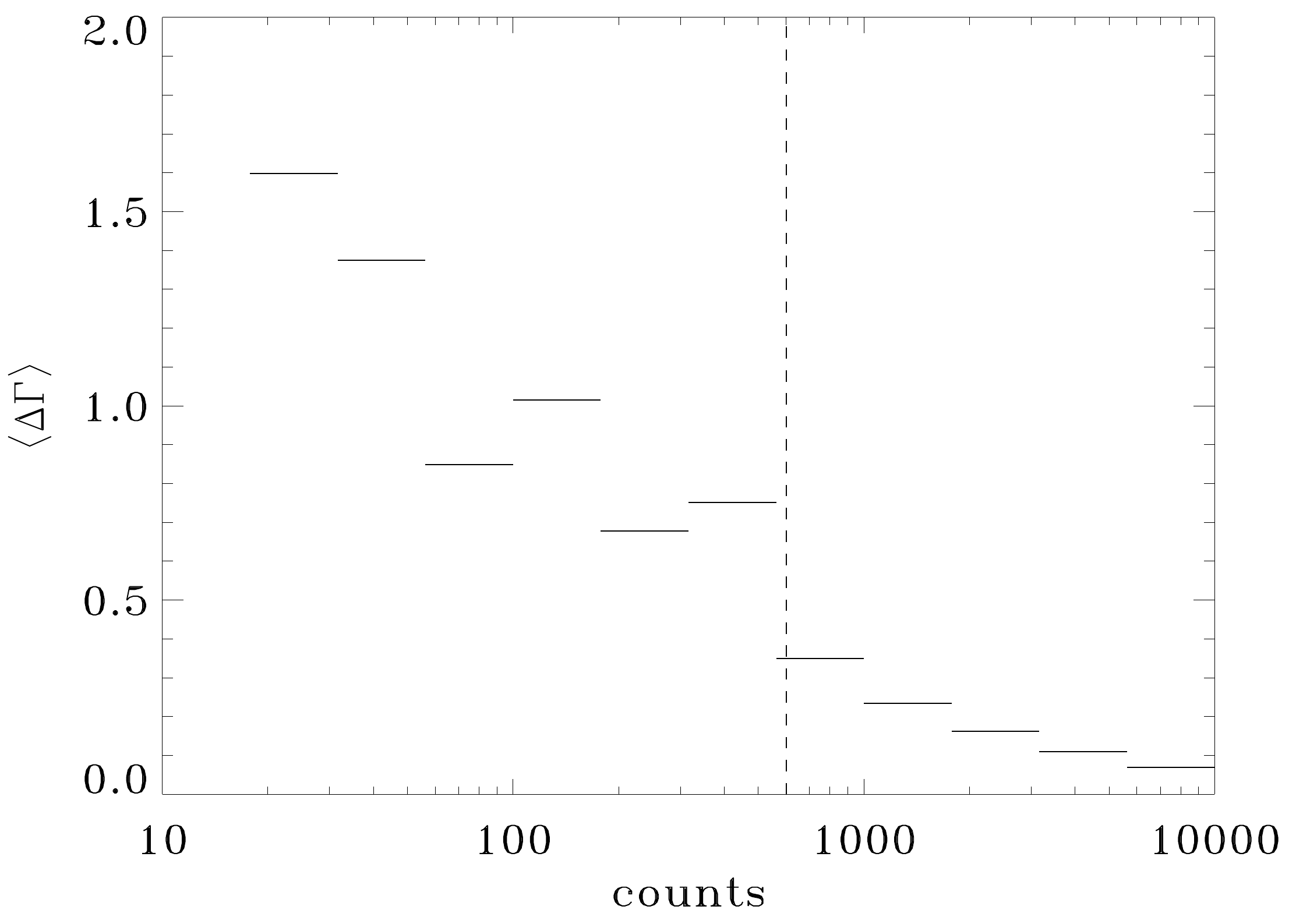}

\caption{The average uncertainty on $\Gamma$ as a function of the number of spectral data counts for the sources in our sample. The dotted line marks 600 counts, the minimum number of counts we require to leave $\Gamma$ free for fitting}
 \label{gammaunc}
\end{figure}

For spectra with a low number of counts, it can be difficult to obtain a converged spectral fit due to large data uncertainties. We perform error analyses on each model fit, using the {\sc xspec} {\tt error} command, for two interesting parameters and $\Delta$c-stat=4.61, which corresponds to 90\% confidence range.  Despite these actions, it is still difficult to constrain the fit parameters well with low count statistics. Fig \ref{gammaunc} shows the binned average uncertainty on $\Gamma$ as a function of total 0.5-8 keV source counts. Below $\sim600$ counts, $<\Delta\Gamma> > 0.6$, which could lead to higher uncertainties in the other parameters. For this reason, we choose to fix $\Gamma$ at 1.9, the canonical value for Seyfert galaxies \citep{nandra94}, for spectra with less than 600 counts. However, in this case, if the best fit model for which $\Gamma$ remains free provides a significantly better fit at the 90\% confidence level ($\Delta$c-stat$>2.71$) than for the best fit model where $\Gamma$ is fixed at 1.9, we select the best fit model where $\Gamma$ remains free. The motivation behind including this additional criterion for allowing $\Gamma$ to be free when there is less than 600 counts is to account for sources with extreme $\Gamma$ values. These include for example, Blazars, which can present very high ($\sim3$) to very low ($\sim1$) $\Gamma$ values \citep{donato05}. Without this consideration, these sources may be identified as Compton thick if $\Gamma$ were fixed to 1.9, due to their flat spectra. This is done for 60/310 sources with less than 600 counts. However, more than half of these have more than 200 counts, where $\Gamma$ can still be fairly well constrained.

\subsection{Identifying Compton thick sources}

The second major aim of this work is to identify Compton thick sources in the CDFS  using these new spectral models. Strictly a Compton thick source is that with \nh$>1.5\times10^{24}$ \cmsq, however, for simplicity here we use $10^{24}$ \cmsq\ as the lower limit. We then aim to assess the Compton thick fraction of sources (sources with $10^{24}<$\nh$<10^{26}$ \cmsq\ with respect to those with $10^{20}<$\nh$<10^{26}$ \cmsq) at high redshift. Spectral fitting with models which correctly account for Compton scattering and the geometry of the absorber and which also include self consistent Fe K$\alpha$ emission are valuable for identifying Compton thick sources, as the equivalent width of the Fe K$\alpha$ emission is tightly linked to \nh\ \citep{leahy93,ghisellini94,brightman11}. Furthermore, intrinsic source luminosity calculations are more realistic when using these new models. Models based on a slab geometry such as {\tt pexrav} underestimate the intrinsic source luminosity by several factors when fitting Compton thick sources \citep{murphy09, brightman11}.  This is due to the fact that {\tt pexrav} assumes a slab of optically thick material that subtends a solid angle of 2$\pi$ steradians from the perspective of a point source, relevant for an optically thick accretion disk. However, the torus is likely to subtend a much smaller solid angle, and thus the reflection strength is weaker with respect to the accretion disk. Therefore the intrinsic source luminosity for a torus model is higher than that for {\tt pexrav} for the same reflection strength. For identifying Compton thick sources, we utilise all 449 sources with available redshifts, be it spectroscopic or photometric.

\begin{figure}
\includegraphics[width=90mm]{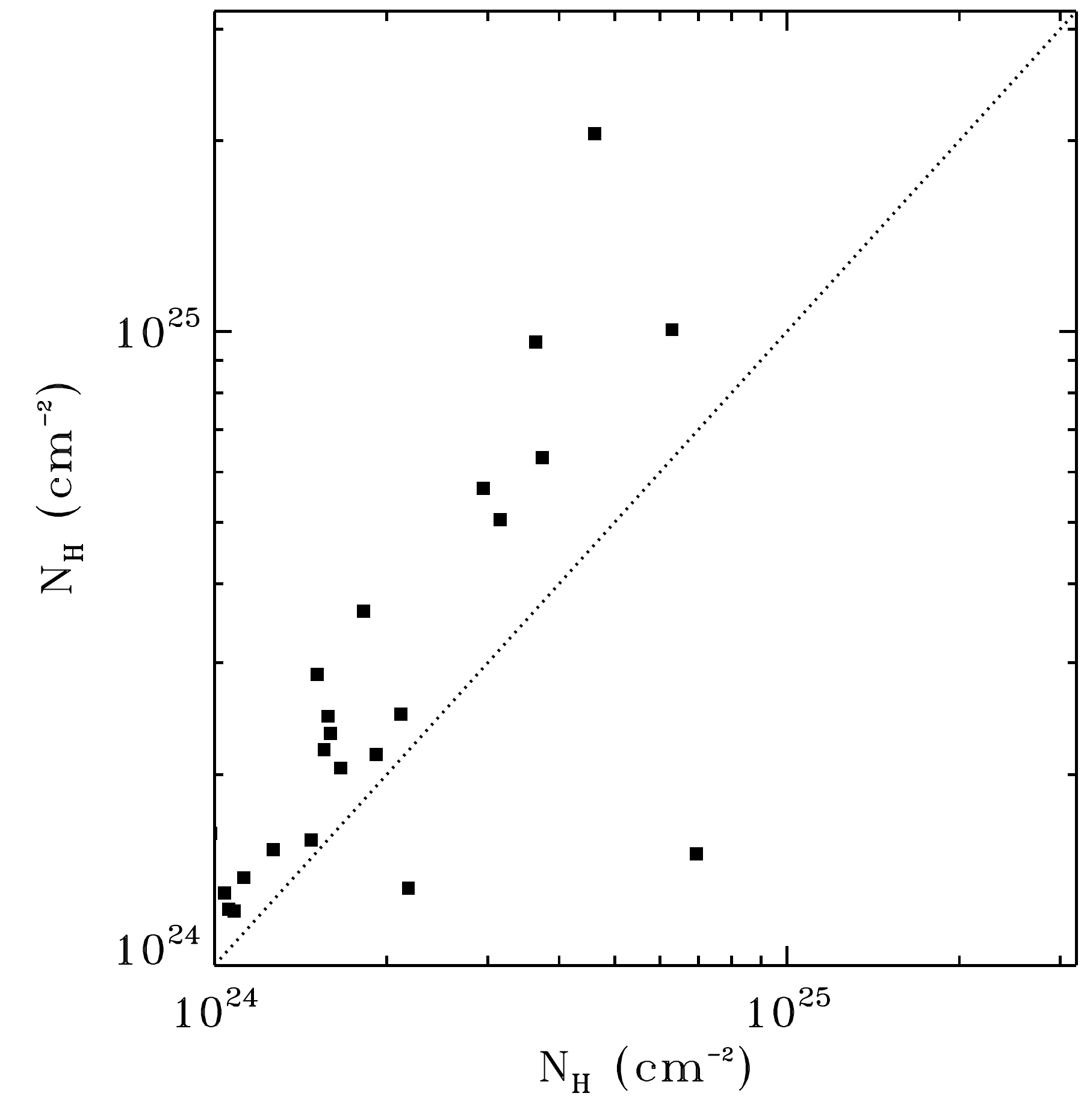}

\caption{The \nh\ as determined by the torus model (x-axis) against the \nh\ as determined by the analytical models, for heavily absorbed sources. It can be seen that the analytical models overestimate the \nh\ with respect to the torus models as they do not account for attenuation of the spectrum by Compton scattering.}
 \label{fig_nhcomp}
\end{figure}

We also fit our spectra with standard analytical models to compare to the results from the torus models. For this purpose we use a simple power-law model, combined with {\tt zwabs} \citep{morrison83} to model photoelectric absorption, {\tt zgauss} (line energy fixed to rest frame 6.4 keV, width fixed to 0.001 keV) to model any Fe K$\alpha$ emission and {\tt pexmon} \citep{nandra07} to model reflection and associated Fe K emission. We also use a secondary power-law component to fit the scattered emission, where the power-law index is fixed to that of the primary power-law component. Once again, for the addition of any free parameters, the c-statistic must improve by 1.0. In Fig \ref{fig_nhcomp}, we show a comparison of the \nh\ as measured by the torus models, with the \nh\ measured by the analytical models for the most absorbed sources. This shows that the analytical models systematically overestimate the \nh\ with respect to the torus models as they do not account for attenuation of the spectrum by Compton scattering.

%Correctly identifying Compton thick sources using X-ray spectroscopy is challenging, especially for spectra with a low number of counts, even with the new torus models. At low signal to noise, a reflection spectrum with a modest scattered component may be mistaken as having no absorption at all, and also vice verse. Therefore, 

\section{results}
\label{results}

\subsection{Geometry of the absorber}

In Table \ref{bfmo_table}, we present the best fit model data, which shows the number of spectra where each model combination has been selected as the best fit model. For low \nh, models A, B and C, which model obscured sight lines, are largely degenerate, as only the geometry of the absorber differentiates them. The geometry has negligible effect below $\sim10^{23}$ \cmsq\ as Compton scattering is not important. In column 3 we show the number of each model selected where the \nh\ is measured to be $>10^{23}$ \cmsq. If we consider the fraction of scattered nuclear X-rays, $f_{\rm scatt}$, which has been shown to be dependant on the opening angle of the torus, we find that heavily absorbed sources with a scattered fraction $f_{\rm scatt}>1.0\%$ are best fit by the torus model with the largest (60\degree) opening angle (model E) over the torus model with the smallest (30\degree) opening angle. Lastly,  we present the mean \fscat\ for the two torus models, for spectral fits where there are at least 600 counts in the spectrum. Model B with the 30\degree\ opening angle gives a systematically smaller \fscat\ ($1.4\pm0.3$\%) than model A with the 60\degree\ opening angle ($2.1\pm0.7$\%). The average \fscat\ we measure for all fits that require the scattered component with $>600$ counts is $1.7\pm0.4$\%. These results present us with a self consistent picture, whereby the fraction of scattered X-rays measured in the spectra is closely linked to the covering factor of the torus. Our results show that while the torus paradigm is still applicable at high redshifts, there is a significant fraction of sources ($\sim20$\%) where the source is likely to be buried in geometrically thick material with close to total  coverage as they are best fit by the 0\degree\ torus model and do not require a scattered component. These sources are analogous to the so called ``new type'' AGN as discovered in the local universe by \cite{ueda07}.

\begin{table*}
\centering
\caption{Details of the best fit models. Column (1) gives the model description; Column (2) gives the number where each model combination is the best fit; Column (3) gives the number where each model combination is the best fit and \nh$>10^{23}$\cmsq; Column (4) gives the number where each model combination is the best fit, \nh$>10^{23}$\cmsq\ and  \fscat\ is $>1$\%; Column (5) is the average \fscat\ for each, where the total number of counts in the spectrum is $>600$}
\label{bfmo_table}
\begin{center}
\begin{tabular}{l r r r r r}
\hline
Model & All & \nh$>10^{23}$\cmsq\ & \nh$>10^{23}$\cmsq\ & $<f_{\rm scatt}>$ (\%)\\
 & & & +$f_{\rm scatt}>1\% $\\
(1) & (2) & (3) & (4) & (5) \\

\hline
(A) 60\degree\ torus			&  77 (19\%) 	& 61 (42\%)	& 50 (34\%) & $2.1\pm0.7$\%\\
(B) 30\degree\ torus			& 83 (20\%) 	& 56 (39\%)	& 40 (28\%) & $1.4\pm0.3$\%\\
(C) 0\degree\ torus (sphere)	& 135 (33\%) 	& 28 (19\%)	& - & -\\
(D) power-law				& 90 (22\%) 	& -			& - & - \\
(E) unobscured torus		& 23 (5.6\%) 	& -			& - & - \\
\hline
Total 					& 408		& 145 		& 90 (62\%) &  $1.7\pm0.4$\\
\hline
\end{tabular}
\end{center}
\end{table*}

\begin{figure}
\includegraphics[width=90mm]{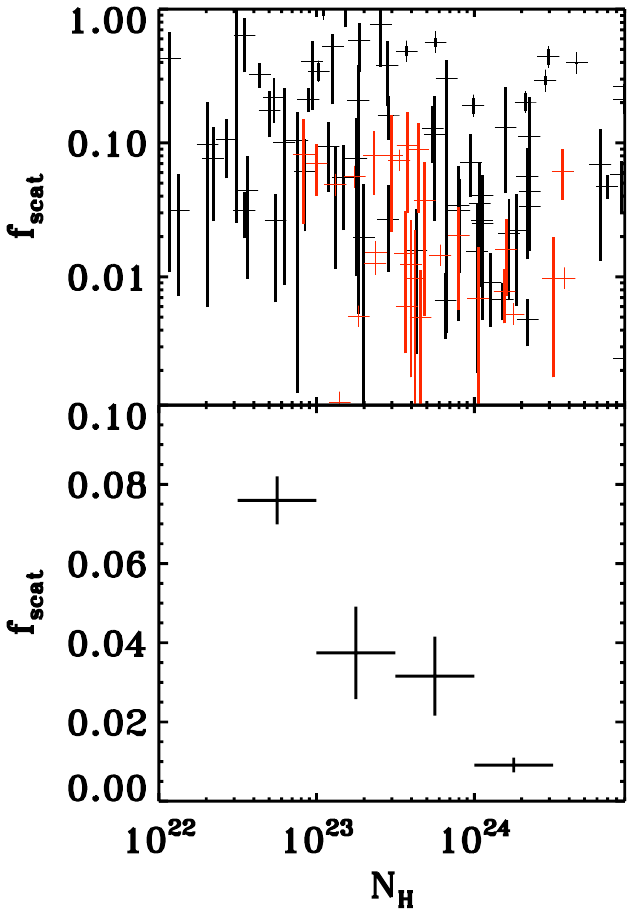}
 \caption{Upper panel: The vertical bars represent the 90\% confidence limit on the range of \fscat\ for each source, against \nh. Red data points are for sources with $10^{43}<$\lx$<10^{44}$ \ergs\ and $1<z<3$ in order to break any degeneracy with those parameters. Lower panels: The binned average of \fscat\ against \nh\ for the subsample defined above (red bars in the upper panel). The error bars represent the standard error (i.e. the standard deviation of the data divided by the square root of the number of data points) on the \fscat\ values.}
 \label{fig_fscat}
\end{figure}

Having found that our results are consistent with \fscat\ being linked to the opening angle of the torus, we can investigate how the covering factor varies with other source parameters  as well as how the fraction of fits which prefer a torus with high covering factor over a lower covering factor behaves. In Fig. \ref{fig_fscat}, we plot \fscat\ against \nh. We plot as thick red bars only sources with $10^{43}<$\lx$<10^{44}$ \ergs\ and $1<z<3$ in order to break any degeneracy with these parameters. We do not explore the relationship between \fscat\ and \lx\ as there may be contamination of scattered X-rays from star forming processes in the host galaxy, the relative strength of which are likely to be linked to the luminosity of the AGN. Furthermore the constraint on \fscat\ decreases towards high redshift  due to the scattered component being redshifted out of the \chandra\ band, and for this reason we do not explore the relationship between \fscat\ and redshift.

 Spectra where \fscat$>10$\% are likely to represent some scenario where the source is only partially covered, or where the \nh\ has changed during the 10 years of observations \citep[e.g. NGC 1365,][]{risaliti09}, and thus not relevant to the scattered X-rays we are considering here. We therefore disregard sources with \fscat$>10$\% in our analysis of the scattered fraction. We plot the binned average of \fscat\ for the various sub-samples in the lower panels. We find that \fscat\ is strongly dependent on \nh, with \fscat\ decreasing with increasing \nh, implying that the covering factor of the torus increases with line of sight obscuration.
 
To investigate the effect of spectral variability in more detail, we look for photometric variablity in the most obscured sources (\nh$>10^{23}$ \cmsq), which is likely to be caused by spectral variations rather than changes in the intrinsic source power. During spectral extraction, the software package {\sc acis extract} computes the significance of the Kolmogorov-Smirnov (KS) statistic when comparing a uniform count rate model to the observed distribution of source event time stamps. For heavily absorbed sources where this KS statistic is $<0.1$, suggestive of variability, we split the observation into two, analysing the observations made in 2000 and 2007 (2Ms) and the observations made in 2010 (2Ms). For 22 sources where the \nh\ has changed by $\Delta$\nh$>$1 dex, from the first 2Ms to the second, we remove them from our analysis of the scattered fraction. We find that removing these sources has no effect on our conclusions regarding the dependance of \fscat\ on \nh.

\begin{figure*}
\includegraphics[width=180mm]{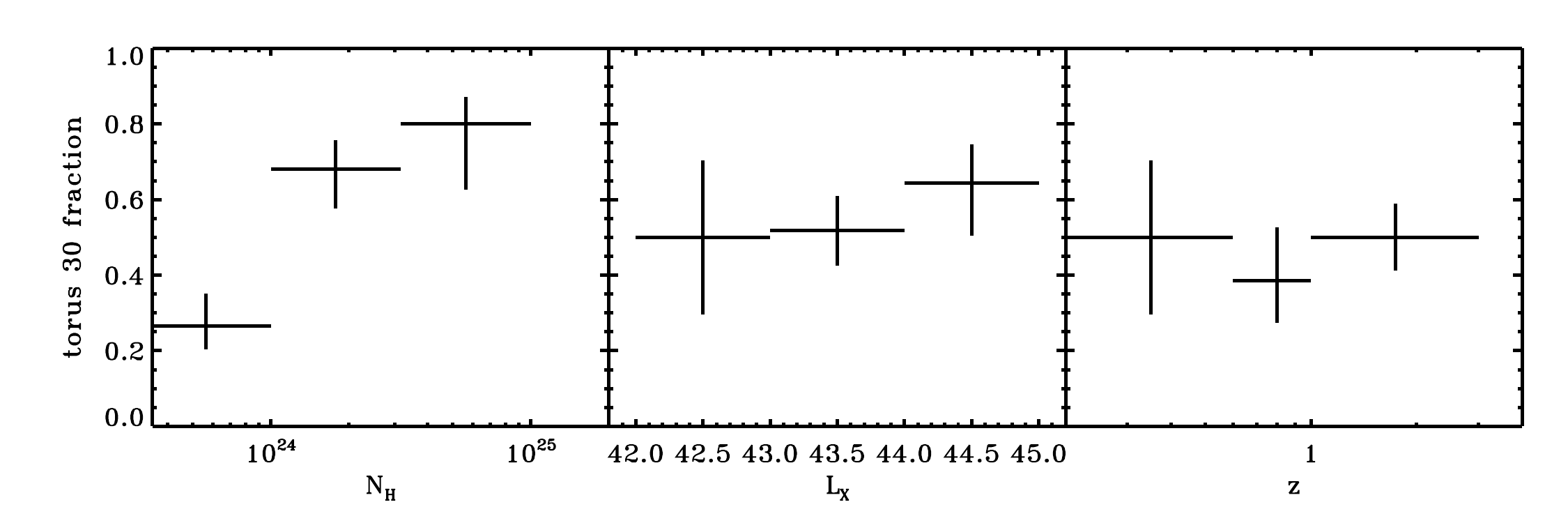}

\caption{The fraction of sources where the torus model with 30\degree\ opening angle is chosen as best fit model over the torus model with 60\degree\ opening, against \nh\ and \lx\ and redshift. We have plotted only for sources with \nh$>10^{23.5}$ \cmsq, as at low \nh\ values these models are degenerate as Compton scattering is not significant. Against \lx\ we plot only for $1<z<3$, and against redshift we plot only for $10^{42}<$\lx$<10^{44}$ \ergs}
 \label{fig_t30f}
\end{figure*}

We also investigate the relative prevalence of the two torus models with different opening angles, and how this varies with \nh, \lx\ and redshift. In Fig \ref{fig_t30f}, we plot the fraction of sources for which the torus model with a 30\degree\ opening angle is preferred as best fit over the torus model with a 60\degree\ opening angle for sources with \nh$>10^{23.5}$ \cmsq, as at low \nh\ values these models are degenerate as Compton scattering is not significant. Again, in order to break degeneracies, for \lx\ we plot only for $1<z<3$, and against redshift we plot only for $10^{42}<$\lx$<10^{44}$ \ergs. These results also indicate that the most heavily obscured sources are in fact the most heavily buried, as the torus model with the smallest opening angle is preferred for the most heavily obscured sources, whereas the torus model with the largest opening angle is preferred in the least obscured sources. This supports our conclusion from the \fscat\ data.

\subsection{Identifying Compton thick sources}
\label{results_ctfrac}

\begin{figure*}
\includegraphics[width=180mm]{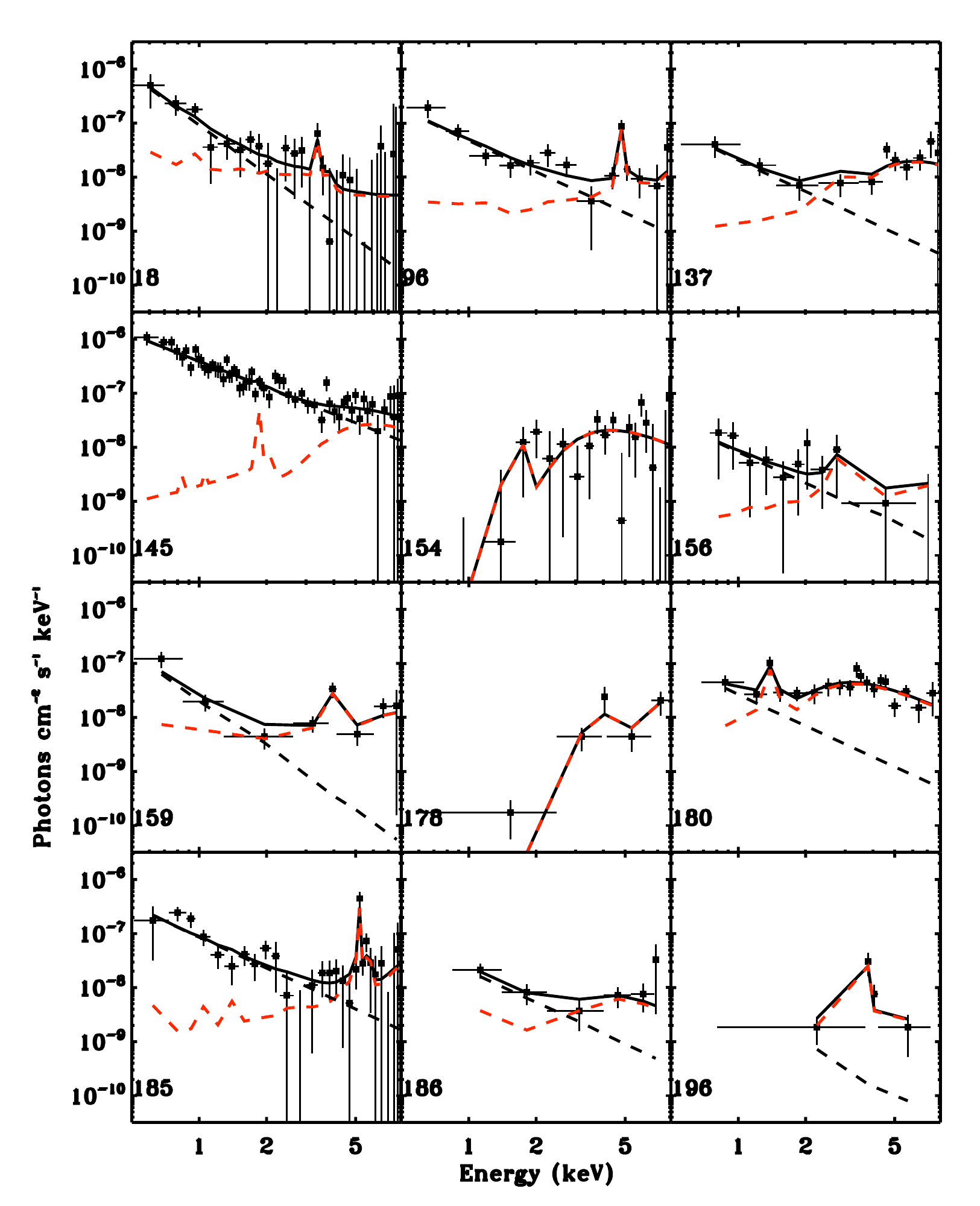}
 \caption{Unfolded spectra of the secure Compton thick sources found in this analysis, using the torus models of BN11. The red dashed lines show the torus component itself, which includes self consistent Fe K$\alpha$ emission, whereas the black dashed lines show the scattered nuclear component, which typically accounts for $<1\%$ of the direct emission in these sources.}
 \label{fig_ctspec1}
\end{figure*}

\begin{figure*}
\includegraphics[width=180mm]{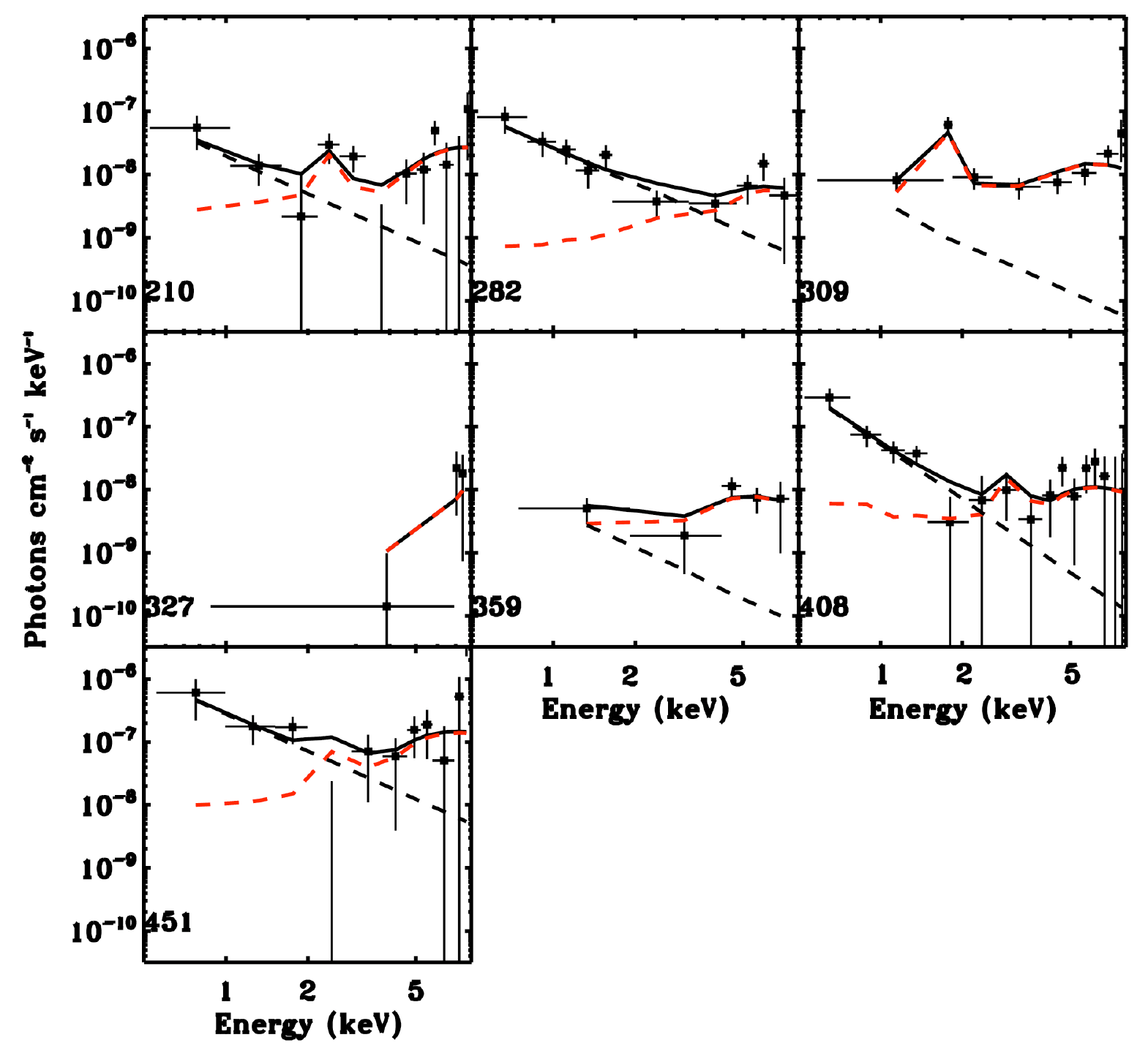}
 \caption{Unfolded spectra of the secure Compton thick sources found in this analysis, using the torus models of BN11. The red dashed lines show the torus component itself, which includes self consistent Fe K$\alpha$ emission, whereas the black dashed lines show the scattered nuclear component, which typically accounts for $<1\%$ of the direct emission in these sources.}
 \label{fig_ctspec2}
\end{figure*}

\begin{figure*}
\includegraphics[width=120mm]{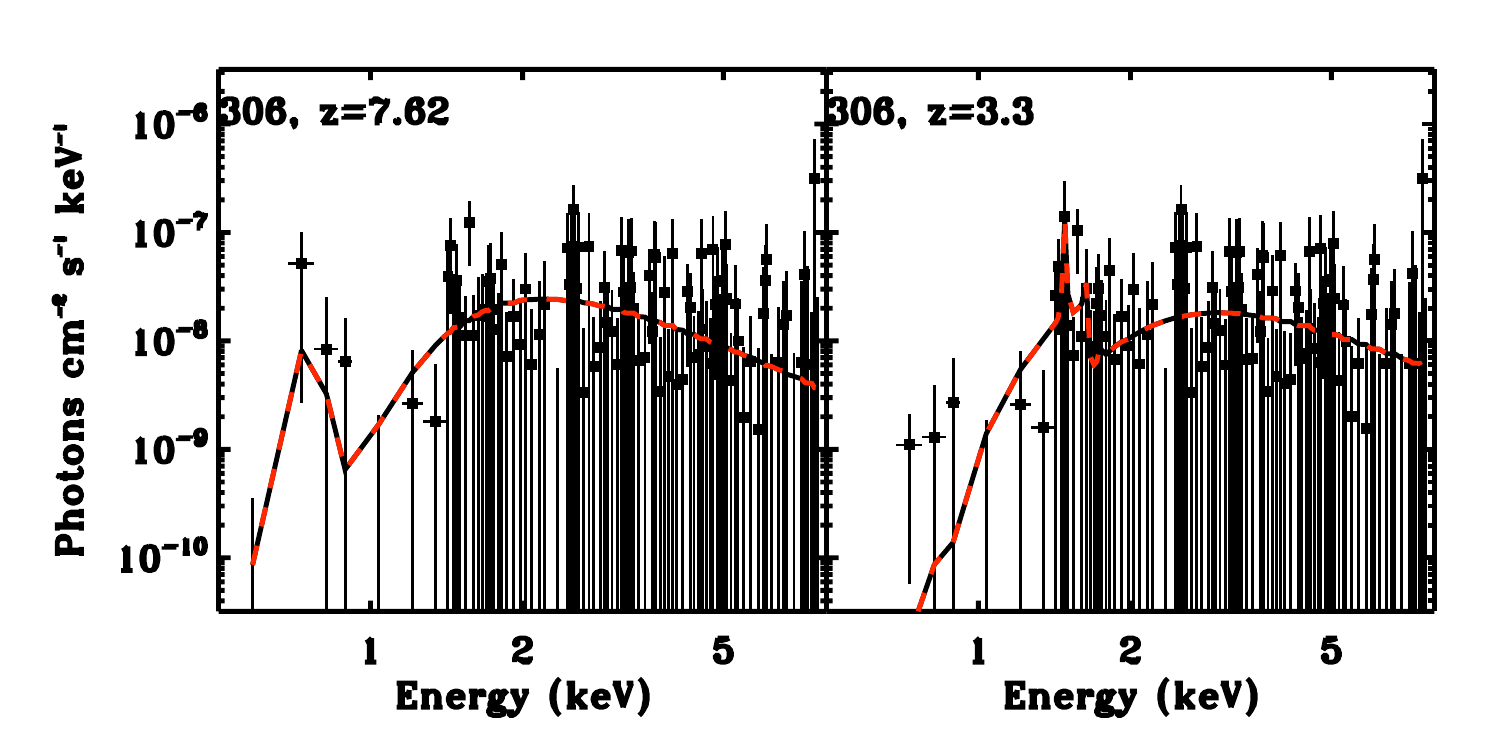}
 \caption{Unfolded spectrum of source 306, found to be Compton thick in our analysis, given a photometric redshift of 7.62 by Luo, et al (2010). We have not binned the spectrum for plotting here so as to show the line at $\sim0.7$ keV, which is consistent with Fe K$\alpha$ emission at the photometric redshift. The spectrum of this source at the alternative photometric redshift given by Luo, et al (2010) of 3.3 is also shown for comparison.}
 \label{fig_ctspec4}
\end{figure*}

Of the 449 sources we study here, we find that 41 of them are Compton thick from our analysis, giving an observed Compton thick fraction for all sources of 9\%. We present the details of the spectral fit parameters, including error analyses, of 20 of these sources which have been found at the 90\% confidence limit ($\Delta$c-stat criterion of 2.71), and are thus the most secure. These are presented in Table \ref{ct_table},  and  the spectra are presented in Figs \ref{fig_ctspec1}, \ref{fig_ctspec2} and \ref{fig_ctspec4}. These Compton thick sources are predominantly best fit by the torus model with the 30\degree\ opening angle, and have very small \fscat\ values (11 fits with \fscat\ $<1.0\%$), which supports our conclusion that the most heavily absorbed sources are in fact the most geometrically buried as well. Due to the low count nature of these sources, the power-law index has been frozen to 1.9 in all but four fits. A total of nine of the Compton thick sources have intrinsic \lx$>10^{44}$ \ergs, making them type 2 quasars.

The highest redshift Compton thick source we find is at $z=7.62$ (306). The redshift is photometric and given by \cite{luo10}. However, an alternative photometric redshift of 3.3 is also presented by the authors. If we allow the redshift in the spectral fit to be a free parameter, we find that the best fit redshift is in fact 6.9$^{+2.9}_{-1.4}$, which is consistent with the preferred  photometric redshift  in \cite{luo10}. If the source were indeed at z=7.62, this  would make this source the highest redshift Compton thick AGN to date and with an intrinsic deabsorbed 2-8 keV luminosity of $3.0\times10^{44}$ \ergs. We present the spectral fit details and spectra for both photometric redshifts given in \cite{luo10}.

\begin{table*}
\centering
\caption{Details of the secure Compton thick sources found in this analysis. Column (1) is the Luo, et al (2008) source number; Column (2) is the number of 0.5-8 keV spectral counts; Column (3) is the redshift of the source; Column (4) is the best fit model for the spectrum as described in section \ref{specanalysis}; Column (5) is the Cash statistic of the fit; Column (6) is the measured \nh\ in units of $10^{24}$ \cmsq; Column (7) is the photon index, $\Gamma$, where $^{\dag}$ indicates that the index was frozen in the fit; Column (8) gives the scattered fraction, \fscat; Column (9) gives the unabsorbed rest-frame 2-10 keV luminosity of the source; Column (10) gives the redshift reference from Table 4 of Silverman, et al (2010). The redshift for source 306 was taken from Luo, et al (2010) and is a photometric redshift.  For this source we have presented spectral fits for the photometric redshift presented (7.62), and the alternative photometric redshift given by the authors (3.3). The following sources are in addition to those presented below and have been detected with 1-$\sigma$ significance -  8, 82, 121, 162, 170, 201, 208, 223, 258, 260, 262, 265, 313, 314, 316, 330, 337, 339, 376, 390 and 399. } 
\label{ct_table}
\begin{center}
\begin{tabular}{r r l c r l l l c c}
\hline
ID & counts & z & model & c-stat & \nh/$10^{24}$ \cmsq\ & $\Gamma$ & \fscat\ (\%) &log$_{10}$(\lx/\ergs) & ref.\\
(1) & (2) & (3) & (4) & (5) & (6) & (7) & (8) &(9) & (10) \\

\hline
18	&	492	&	0.845	&	B	&	174.9	&	8.91	$_{-	8.84	}^{+	91.09	}$ &	3.00	$_{-	2.00	}^{+	0.00	}$ &	0.24	$_{-	0.13	}^{+	0.14	}$ &	44.19 & 	15 \\
96	&	153	&	0.31		&	B	&	102.0	&	2.18	$_{-	1.40	}^{+	97.82	}$ &	1.90	$^{\dag}$ &	0.48	$_{-	0.18	}^{+	0.21	}$ &	42.95 &	1 \\
137	&	111	&	1.544	&	B	&	96.9		&	1.59	$_{-	0.56	}^{+	1.47	}$ &	1.90	$^{\dag}$ &	1.60	$_{-	0.87	}^{+	1.10	}$ &	43.73 &	1 \\
145	&	838	&	2.494	&	B	&	284.5	&	2.95	$_{-	2.02	}^{+	97.05	}$ &	1.62	$_{-	0.30	}^{+	0.36	}$ &	43.95	$_{-	5.75	}^{+	6.18	}$ &	44.12 &	6 \\
154	&	269	&	2.93		&	C	&	169.0	&	1.82	$_{-	0.89	}^{+	2.38	}$ &	1.90	$^{\dag}$ &	$<3.78$ &	44.08 &	8 \\
156	&	30	&	1.413	&	B	&	17.6		&	50.10	$_{-	50.01	}^{+	49.90	}$ &	1.90	$^{\dag}$ &	1.11	$_{-	0.81	}^{+	0.77	}$ &	43.41 &	2 \\
159	&	79	&	0.664	&	B	&	70.3		&	1.58	$_{-	0.66	}^{+	0.65	}$ &	2.87	$_{-	1.14	}^{+	0.13	}$ &	0.13	$_{-	0.07	}^{+	0.10	}$ &	43.53 &	1 \\
178	&	46	&	0.735	&	C	&	43.8		&	1.92	$_{-	0.99	}^{+	2.36	}$ &	1.90	$^{\dag}$ &	$<0.24$ &	43.44 &	1 \\
180	&	208	&	3.66		&	A	&	133.5	&	1.66	$_{-	0.59	}^{+	1.39	}$ &	1.90	$^{\dag}$ &	2.12	$_{-	1.45	}^{+	1.73	}$ &	44.62 &	1 \\
185	&	428	&	0.227	&	B	&	226.9	&	1.51	$_{-	0.83	}^{+	16.92	}$ &	1.90	$^{\dag}$ &	0.68	$_{-	0.20	}^{+	0.23	}$ &	42.76 &	5 \\
186	&	65	&	3.98		&	B	&	68.3		&	6.30	$_{-	5.50	}^{+	93.70	}$ &	1.90	$^{\dag}$ &	6.94	$_{-	5.62	}^{+	5.68	}$ &	44.06 &	8 \\
196	&	27	&	0.668	&	B	&	20.0		&	4.46	$_{-	3.83	}^{+	95.54	}$ &	1.90	$^{\dag}$ &	0.07	$_{-	0.07	}^{+	0.23	}$ &	43.25 &	1 \\
210	&	319	&	1.57		&	B	&	214.6	&	3.16	$_{-	2.66	}^{+	96.84	}$ &	1.90	$^{\dag}$ &	0.66	$_{-	0.66	}^{+	0.81	}$ &	44.10 &	8 \\
282	&	61	&	2.223	&	B	&	56.8		&	3.64	$_{-	2.43	}^{+	96.36	}$ &	1.90	$^{\dag}$ &	6.10	$_{-	2.34	}^{+	2.94	}$ &	43.64 &	1 \\
306 (1)  &       108	&	7.62	         &       C	&	68.5		&	2.49 $_{-  0.91  }^{+  1.31           }$ &  1.90 $^{\dag}$ &       $<8.73$ & 44.47 & X \\
306  (2) &       108	&	3.3	         &       C	&	68.8		&	0.94 $_{-  0.31  }^{+  0.39           }$ &  1.90 $^{\dag}$ &       $<1.12$ & 43.79 & X \\
309	&	78	&	2.579	&	B	&	58.4		&	11.91	$_{-	7.21	}^{+	88.09	}$ &	1.90	$^{\dag}$ &	0.15	$_{-	0.15	}^{+	0.58	}$ &	44.45 &	5 \\
327	&	15	&	1.096	&	C	&	8.8		&	4.62	$_{-	4.62	}^{+	95.38	}$ &	1.90	$^{\dag}$ &	$<0.08$ &	43.82 &	2 \\
359	&	46	&	2.93		&	B	&	31.3		&	3.74	$_{-	2.24	}^{+	96.26	}$ &	1.90	$^{\dag}$ &	0.98	$_{-	0.98	}^{+	2.15	}$ &	43.91 &	8 \\
408	&	188	&	1.115	&	B	&	127.7	&	1.55	$_{-	1.01	}^{+	1.20	}$ &	2.90	$_{-	1.07	}^{+	0.10	}$ &	0.77	$_{-	0.32	}^{+	0.37	}$ &	43.84 &	5 \\
451	&	128	&	1.335	&	B	&	97.3	&	1.85	$_{-	1.51	}^{+	6.70	}$ &	1.90	$^{\dag}$ &	2.23	$_{-	1.62	}^{+	1.96	}$ &	44.56 &	6 \\
\hline
\end{tabular}
\end{center}
\end{table*}

\subsection{Evolution of the Compton thick fraction}

Previous works have shown that the fraction of absorbed AGN increases with redshift \citep[e.g. ][]{hasinger08}. The wide redshift range of the sources in the CDFS allows for an investigation into if and how the Compton thick fraction of the AGN population evolves with epoch. We decide to investigate this evolution with three redshift bins, z=0.1-1,1-2.5 and 2.5-4, designed to cover the redshift range of the CDFS. In order to calculate the intrinsic Compton thick fraction of each bin, we must take into account several biases and limiting factors in our data. Firstly we must determine the reliability of our spectral fitting method to correctly identify Compton thick sources, especially at the faintest fluxes, which we do via simulations. Secondly, as the sample is X-ray selected, it is biased against the most absorbed sources, due to severe flux suppression in the X-ray band. Furthermore, the survey solid angle is a sharply decreasing function of flux below $F_{\rm 0.5-8}\sim10^{-15}$ erg cm$^{-2}$ s$^{-1}$ \citep{luo08}, which means that it is further biased against faint absorbed sources, for which a correction must be made. The source catalogue of \cite{luo08} is a band merged sample, detected in the soft (0.5-2 keV), hard (2-8 keV) and full (0.5-8 keV) bands. To account for the factors described above, we require a single band selected sample. We thus use the full band selected sample, as it has the greatest number of sources (406).

\subsubsection{Simulations}

\begin{figure}
\includegraphics[width=90mm]{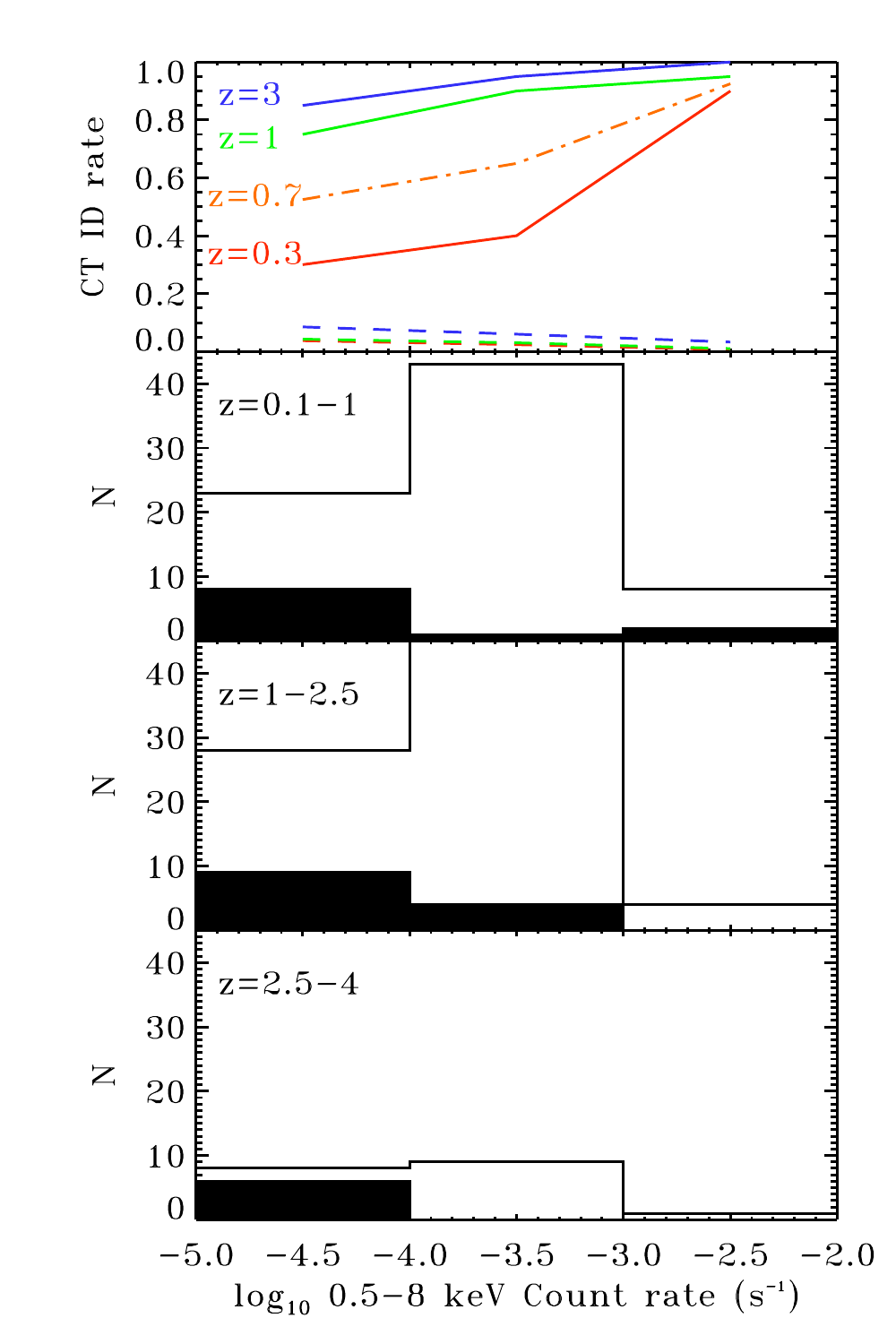}

\caption{Top panel: Solid lines - the rate of correct identification of CT sources from our simulations for a range of count rates, for sources at z=0.3,1.0,3.0. The dot-dashed line is an interpolation of the z=0.3 and z=1 results to z=0.7; dotted lines - the fraction of Compton thin sources incorrectly identified as Compton thick for the same redshifts. Bottom panels: The distribution in 0.5-8 keV count rates for sources at z=0.1-1, 1-2.5 and 2.5-4. The filled histogram shows the Compton thick sources}
 \label{fig_ctid}
\end{figure}

\begin{figure}
\includegraphics[width=90mm]{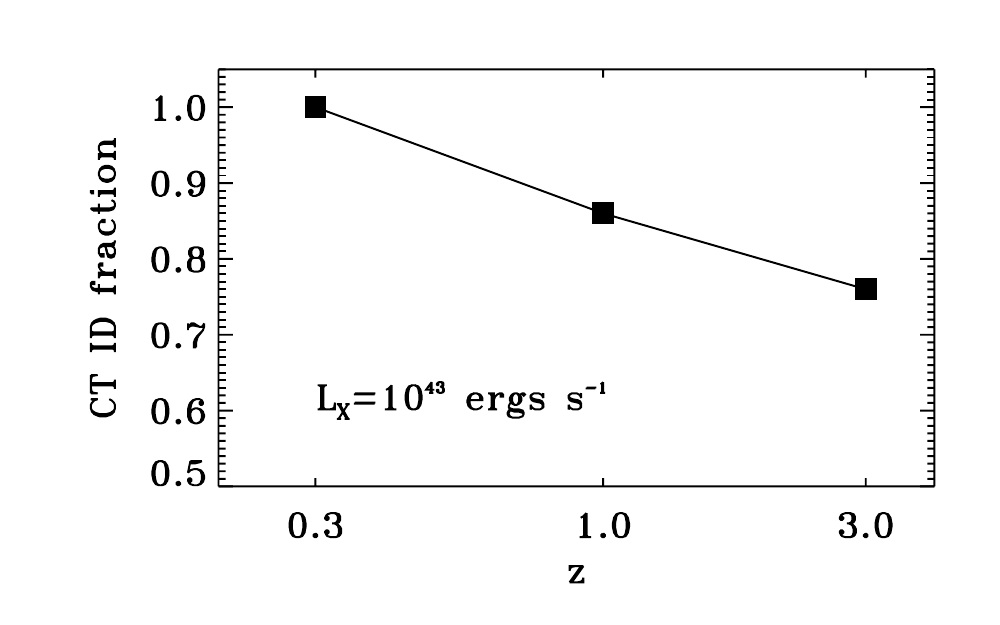}

\caption{The rate of correct identification of CT sources from our simulations for a source with \lx=10$^{43}$ \ergs\ as a function of redshift.}
 \label{fig_ctidz}
\end{figure}

Above we have described the details of the most secure Compton thick sources from our analysis, at the 90\% confidence limit. In our spectral analysis, we adopted the 1-$\sigma$ significance level when adding parameters to the spectral model, such as absorption. We run simulations in order to ascertain the fraction of Compton thick AGN which are correctly identified using our spectral fitting method, as well as to determine the number of contaminating sources, sources which are incorrectly identified as Compton thick. We use the {\tt fakeit} command in {\sc xspec} to produce simulated spectra, assuming a torus model with 30\degree\ opening angle and a 1\% scattered component, using the \chandra\ response and a typical background file from a typical source in the CDFS. We simulate 100 spectra for each \nh\ ($10^{20.5}, 10^{21.5}, 10^{22.5}, 10^{23.5}, 10^{24.5}, 10^{25.5}$), for three different 0.5-8 keV count rates ($10^{-4.5}, 10^{-3.5}, 10^{-2.5}$ s$^{-1}$) and three redshifts (z=0.3, 1.0 and 3.0).  The rate of Compton thick identification and the contamination rate determined from these simulations are plotted in Fig. \ref{fig_ctid}, as well as the distribution of count rate in three redshift ranges, corresponding to the simulations. We find that, as expected, the rate of Compton thick identification is at its lowest at $<40$\% at the faintest count rates, but also in the lowest redshift sources. This of course improves with increasing count rate and also as the redshift increases. A significant increase in the rate of identification is seen from redshift 0.3 to 1, due to the soft excess emission being redshifted out of the band pass and the reflected light being redshifted into it. We thus also plot the interpolation of the identification rate for z=0.7. The contamination rate is $<10$\% for all count rates and redshifts. Plotted in Fig. \ref{fig_ctidz} is the rate of CT identification for a source with an intrinsic \lx=10$^{43}$ \ergs, as a function of redshift. This shows that despite the \chandra\ band observing a higher rest-frame energy for higher redshift sources, and thus being less affected by absorption, identification of Compton thick AGN becomes more difficult as redshift increases, for a constant luminosity.

While simulated data can give us statistical information on the rate of Compton thick identification, it is difficult to determine which individual sources have been correctly identified, or not identified at all. One test that can be performed is to check if optical broad lines have been detected in any of the sources identified as Compton thick, as these sources are unlikely to be Compton thick in reality. We find that 9 such sources are optical broad line sources \citep[given in][]{silverman10} and have been identified as Compton thick in our analysis. We therefore remove these from our list of Compton thick candidates. However, this contamination rate is consistent with data from our simulations. Given our knowledge about the rate of Compton thick identification from these simulations and the rate of contamination, we correct on a statistical basis, the observed number of Compton thick sources. For the z=1-2.5 bin we use the z=1 simulation results to correct the numbers, and for the z=2.5-4 bin, we use the z=3 results. As we have shown from our simulations, the rate of Compton thick identification increases significantly from z=0.3 to z=1. The redshift distribution of our z=0.1-1 sources peak around z=0.7, and thus to avoid underestimating the identification rate, we use the interpolation of the z=0.3 and z=1 results to correct the numbers in this bin.

\subsubsection{Correcting for survey biases}

\begin{figure}
\includegraphics[width=90mm]{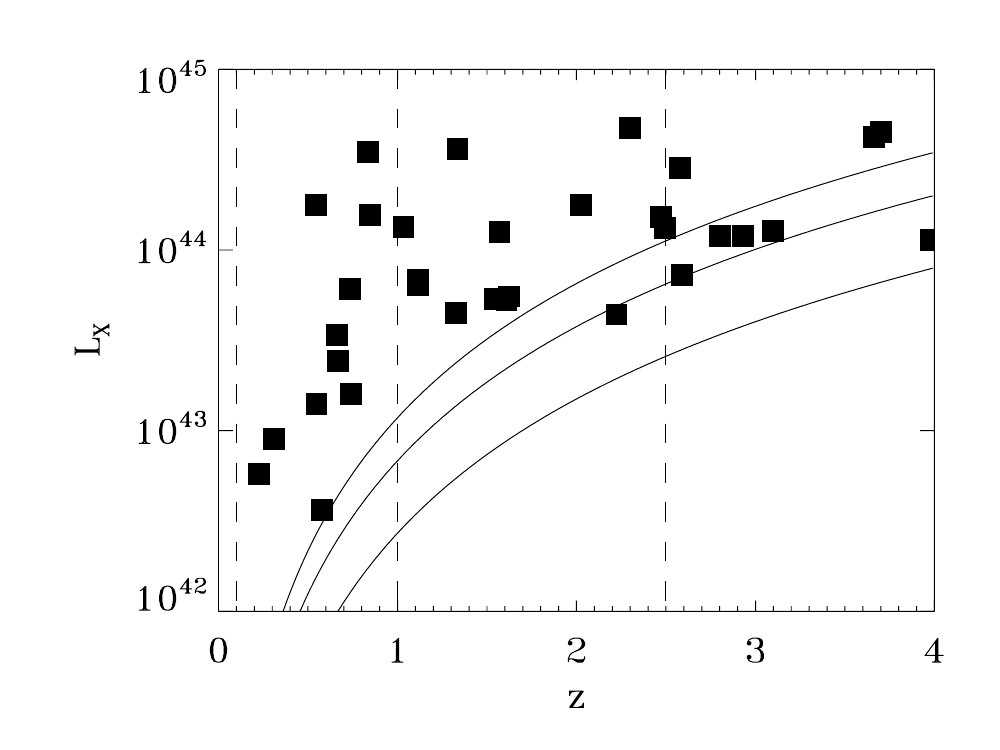}

\caption{The intrinsic 2-10 keV luminosity of the CTAGN with a 0.5-8 keV count rate of $>10^{-5}$ counts s$^{-1}$ in our sample against redshift. The vertical dashed lines show the redshift binning we use, z=0.1-1,1-2.5,2.5-4, while the solid curve represents the intrinsic luminosity we are sensitive to sources up to \nh$=10^{24}, 10^{24.5}, 10^{25}$ \cmsq\ with the adopted count rate limit.}
 \label{fig_ctagnzlx}
\end{figure}

As our sample is X-ray selected, it is intrinsically biased against heavily obscured AGN, due to the significant flux suppression occurring in these sources. In order to account for this bias, we restrict our analysis to an intrinsic luminosity range, where the lower limit is determined by the intrinsic luminosity to which the survey is sensitive to Compton thick sources. Below $10^{-5}$ counts s$^{-1}$ in the 0.5-8 keV band, the survey solid angle becomes very small, which could introduce large uncertainties in our data due to small number statistics. We therefore apply a count rate cut at this level to our sample. Fig. \ref{fig_ctagnzlx} plots the intrinsic 2-10 keV luminosity against redsift of the Compton thick AGN above this count rate limit, which has been determined from the spectral fit model. The solid curves on this plot shows the intrinsic luminosity we are sensitive to sources up to \nh$=10^{24}, 10^{24.5}, 10^{25}$ \cmsq\ with this adopted count rate limit. We then calculate the intrinsic Compton thick fraction by restricting our analysis to sources with intrinsic luminosities above the curve for \nh$=10^{25}$ \cmsq,and \lx$>10^{42}$ \ergs\ in order not to include star forming galaxies. While using this curve we loose a few CTAGN in the highest redshift bins, we can be certain that we are complete to all CTAGN.

To correct for the differential survey solid angle as a function of flux, we bin up our sample into three bins of 1 dex in 0.5-8 keV count rate, and convert the observed number of sources into number densities using the average survey solid angle for the 0.5-8 keV band within that bin. The solid angle data presented in \cite{luo08} is given as a function of flux, which has been converted from a count rate with a power-law of $\Gamma=1.4$, and so we convert this flux back into a count rate using the same model, so that we may carry out the integration. As mentioned above, we apply a count rate cut of $10^{-5}$ counts s$^{-1}$ due to the very small survey solid angle below this limit.

\begin{figure}
\includegraphics[width=90mm]{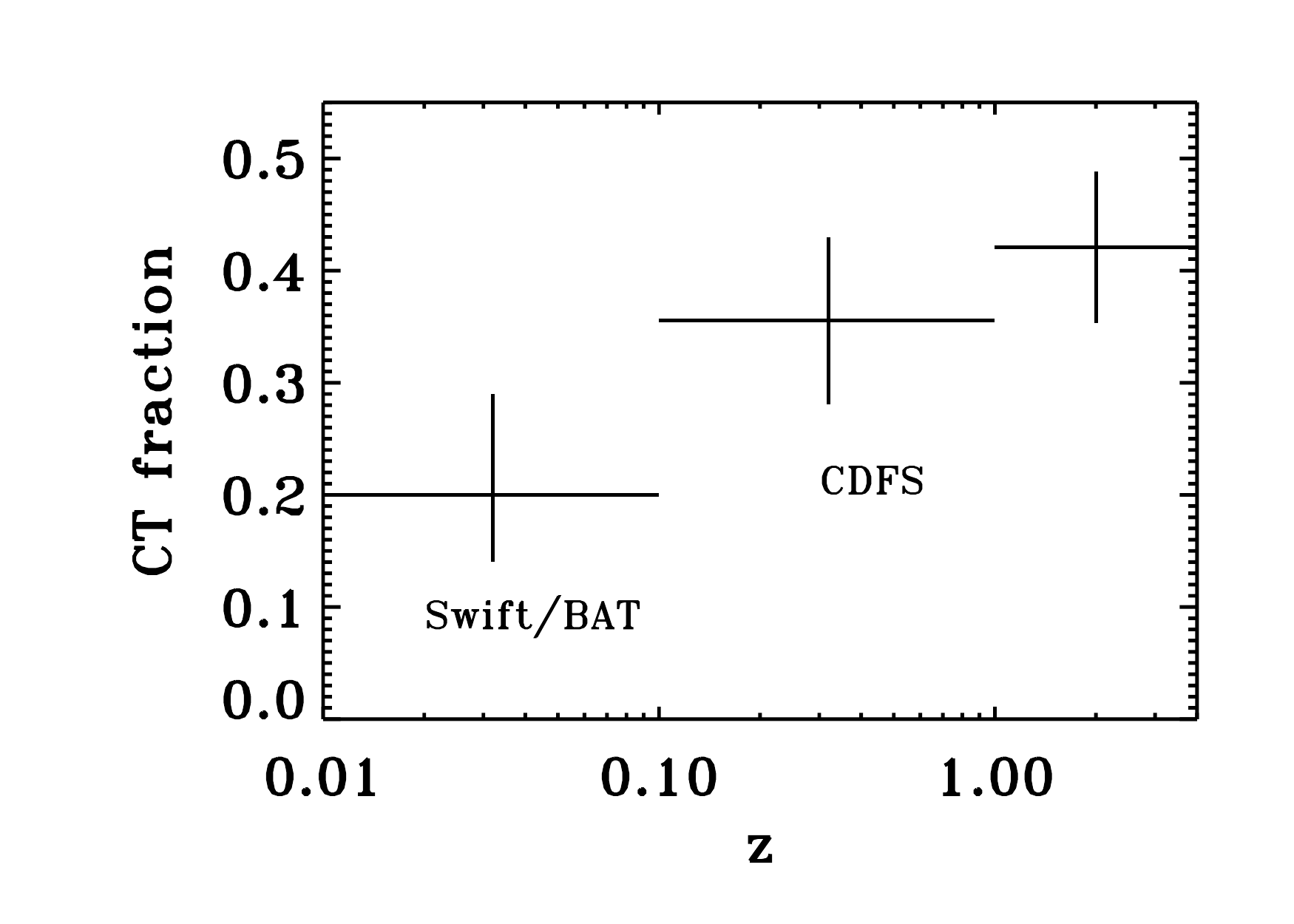}

\caption{The evolution of the Compton thick fraction with redshift, which has been normalised at log$_{10}$\lx=43.5 in each redshift bin, using data from H08. The 1-2.5 and 2.5-4 redshift bins have been merged here to improve the statistics. The uncertainties on $f_{\rm CT}$ have been propagated from the Poisson errors on $N($\nh$>10^{24}$ \cmsq) and $N($\nh$<10^{24}$ \cmsq) The {\it Swift/BAT} sample has an average luminosity of log$_{10}$\lx=43.6 \citep{burlon11}, and so we plot the data from this sample with no renormalisation.}
\label{fig_ctz}
\end{figure}

Once we have calculated number densities, having corrected for all biases aforementioned, we can then calculate the Compton thick fraction for each redshift bin. For each redshift bin a different range in luminosity has been used, due to the increasing luminosity sensitivity to Compton thick AGN with redshift. However, X-ray obscuration is heavily dependent on X-ray luminosity, so in order to evaluate Compton thick fractions as a function of redshift, we must determine the Compton thick fraction at a common luminosity. This was done similarly by \cite{hasinger08}, H08 henceforth, who evaluated the dependence of AGN obscuration on luminosity and redshift. The author finds that $f_{\rm2}$ is a linear function of log$_{10}$\lx\ with a slope of -0.28, where the constant increases with redshift. We require a relation for $f_{\rm CT}$ vs. \lx\ which we convert from $f_{\rm 2}$ vs. \lx\ . In H08, X-ray absorption was determined through the use of hardness ratios, such that $f_{\rm 2}=N(21.5<$log$_{10}$\nh$<24$)/$N(20<$log$_{10}$\nh$<24$). We assume that most of that sample is Compton thin and does not contain many sources with \nh$>10^{24}$ \cmsq.  In order to convert the relation for $f_{\rm 2}$ into a relation for $f_{\rm CT}$, where $f_{\rm CT}=N(24<$log$_{10}$\nh$<26$)/$N(20<$log$_{10}$\nh$<26$), we assume that $N(24<$log$_{10}$\nh$<26$)=$N(21.5<$log$_{10}$\nh$<24$), which is an almost flat \nh\ distribution above \nh=$10^{21.5}$ \cmsq. This gives us the relation $f_{\rm CT}=f_{\rm 2}/(1+f_{\rm 2})$. For each redshift bin, we calculate the average log$_{10}$ \lx, and renormalise $f_{\rm CT}$ to log$_{10}$\lx=43.5 in each bin. We present the data for these three redshift bins in Table \ref{ctz_table}. We plot the normalised Compton thick fraction as a function of redshift in Fig. \ref{fig_ctz}, where we have merged the 1-2.5 and 2.5-4 redshift bins into one in order to improve the statistics, using a simple average. We also plot the results from \cite{burlon11} for the {\it Swift/BAT} sample of low redshift sources. The mean 2-8 keV luminosity for this sample is log$_{10}$\lx/\ergs=43.6, once we have converted the 15-55 keV luminosity using a $\Gamma=1.9$ power-law, and is thus also comparable to the luminosity considered in our analysis. Our data point toward an increasing intrinsic CT fraction from $\approx$20\% at z$<0.1$ to $\approx$40\% at z$>1$ for \lx$=10^{43.5}$ \ergs\ AGN.

\begin{table*}
\centering
\caption{Details of the Compton thick fraction for AGN in the redshift range given by column (1) with an average 2-10 keV log$_{10}$\lx\ for all AGN given in column (2); Column (3) gives the total number of sources for each bin; Column (4) gives the number of \nh$>10^{24}$ \cmsq\ AGN in each bin identified from spectral fitting; Column (5) gives the number of \nh$>10^{24}$ \cmsq\ AGN in each bin after correcting for misidentification and contamination; Column (6) and (7) give the number density of AGN and CTAGN given the survey solid angle; Column (8) gives the Compton thick fraction in the selected luminosity range; Column (9) gives Compton thick fraction normalised at log$_{10}$\lx=43.5. The uncertainties presented are calculated from Poisson uncertainties. The uncertainties on $f_{\rm CT}$ have been propagated from the Poisson errors on $N($\nh$>10^{24}$ \cmsq) and $N($\nh$<10^{24}$ \cmsq)}
\label{ctz_table}
\begin{center}
\begin{tabular}{l c c c c c c c c}
\hline
Redshift			& $<$\lx$>$	& $N_{\rm tot}$	& \multicolumn{2}{c}{$N_{\rm CT}$} 	& $N_{\rm tot}$ & $N_{\rm CT}$ & $f_{\rm CT}$ (\%) &  $f_{\rm CT}$ (\%) \\
				& 			& 			& Ident.	& Cor. 				& (arcmin$^{-2}$)& (arcmin$^{-2}$) & &log$_{10}$\lx=43.5 \\
(1) & (2) & (3) & (4) & (5) & (6) & (7) & (8) & (9)\\
\hline
$0.1\leq z<1.0$ &       43.2&          60&          11&      18.0&     0.176$\pm$    0.022&    0.068$\pm$    0.016&      38.9$\pm$      6.7&       35.5$\pm$  7.5\\
$1.0\leq z<2.5$ &       43.9&          66&          12&      13.8&     0.194$\pm$    0.024&    0.053$\pm$    0.014&      27.1$\pm$      6.0&       32.0$\pm$ 5.2\\
$2.5\leq z<4.0$ &       44.4&          11&           3&      3.1&    0.032$\pm$   0.010&    0.013$\pm$   0.008&      42.2$\pm$      16.3&       49.5$\pm$12.4\\
\hline
\end{tabular}
\end{center}
\end{table*}

\begin{figure}
\includegraphics[width=90mm]{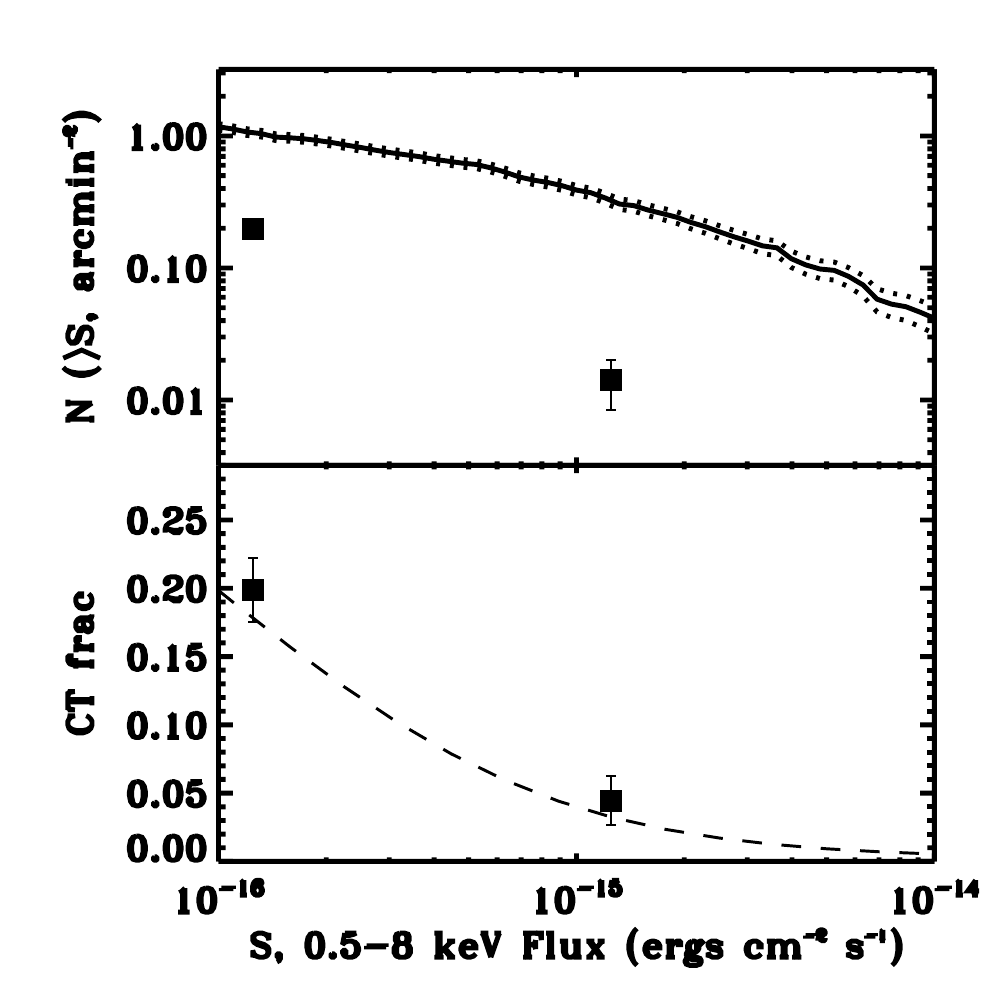}

\caption{Top panel: log N - log S data for AGN (solid line, where the dotted lines show the Poisson upper and lower uncertainties) and CTAGN (data points) with \lx$>10^{42}$ \ergs\ in the full band selected sample of Luo, et al (2008). Bottom panel: The Compton thick fraction as a function of limiting flux in the 0.5-8 keV band. Fluxes have been converted from the 0.5-8 keV count rate using a $\Gamma=1.4$ power-law. The dashed line shows the model prediction from Gilli, et al (2007).}
\label{fig_ctfx}
\end{figure}

We also present number count data as a function of limiting flux (log N - log S) for the full band selected sample of \cite{luo08} for AGN and CTAGN with \lx$>10^{42}$ \ergs. These have been calculated as a function of 0.5-8 keV count rate, and then converted into a flux assuming a power-law model with $\Gamma=1.4$ for convenience. The CTAGN numbers  have been corrected for spectral identification incompleteness and contamination as described previously, but the luminosity ranges have not been restricted as above, only to \lx$>10^{42}$ \ergs.  Using these data, we have calculated the CT fraction as a function of limiting flux, which is very useful for comparing to XRB synthesis models. These data are presented in Fig. \ref{fig_ctfx}, along with the model prediction of \cite{gilli07}. This fluxes from this prediction have been converted from a 2-10 keV flux to a 0.5-8 keV flux using a power-law model with $\Gamma=1.4$. While this comparison is not straightforward, due to differing spectral models used, the observational and model data agree very well. However, the redshift evolution of the absorbed AGN fraction that we find here is not included in the XRB models of \cite{gilli07}, so the agreement we find with their models is likely to be coincidental.

\subsubsection{Alternative Compton thick selection methods}
 
\begin{figure}
\includegraphics[width=90mm]{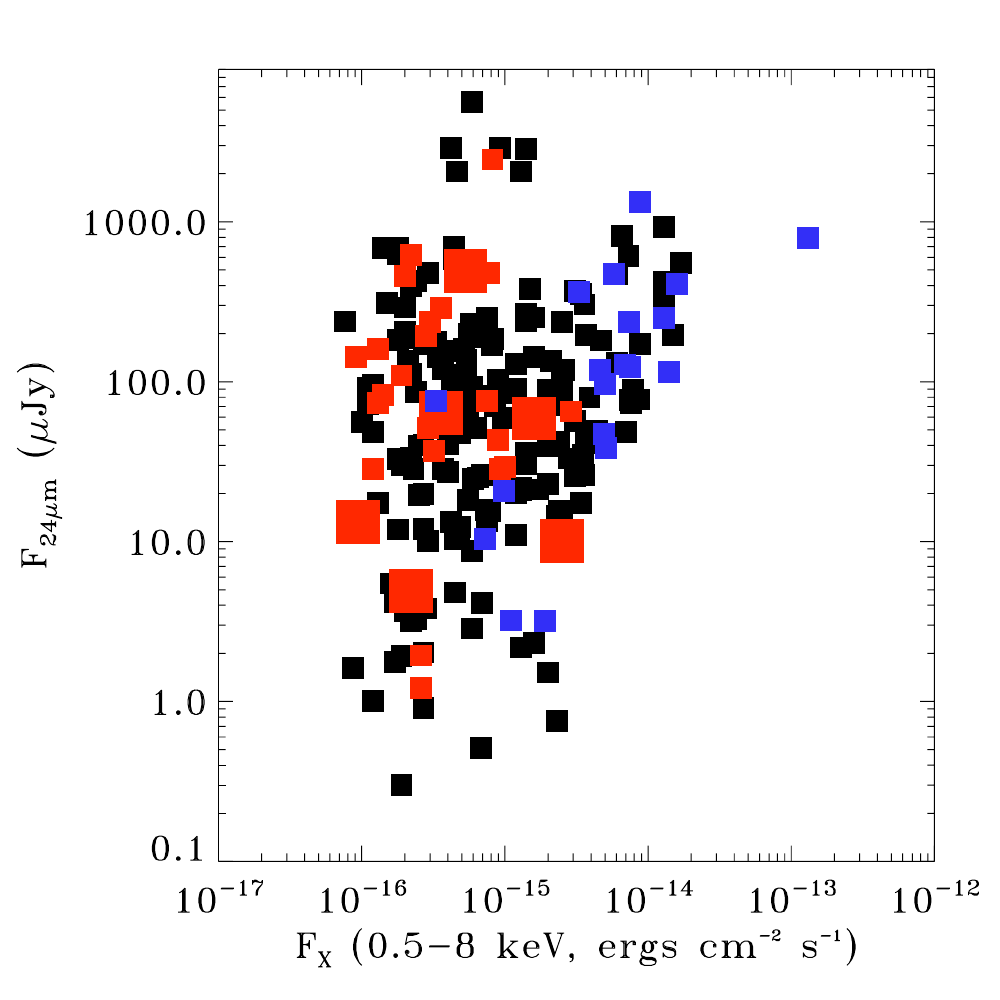}

\caption{The 0.5-8 keV X-ray flux against 24 micron flux for all X-ray sources with a 24 micron counterpart. Blue squares are optical broad line AGN, and red squares are the Compton thick AGN we find here.  The largest red squares are Compton thick quasars with an intrinsic \lx$>10^{44}$ \ergs.}
 \label{fig_fx24}
\end{figure}

A possible alternative method of identifying Compton thick sources, which as we have shown, can be very challenging from X-ray spectral analysis at low fluxes, is to use data from other wavelengths. Most studies have concentrated on mid-infrared (MIR) wavelengths, where the primary radiation absorbed by the circum-nuclear material is reemitted. Here, we briefly explore the multiwavelength properties of our sample starting with the MIR. We obtained 24 micron photometry in the CDFS from the FIREWORKS catalogue of \cite{wuyts08}. We searched for counterparts to the X-ray sources by cross correlating the positions of the X-ray sources with the positions of the 24 micron sources, taking the closest 24 micron source, rejecting any counterpart with an offset of $>5$ arcsecs. Fig. \ref{fig_fx24} plots the X-ray flux against the 24 micron flux of X-ray sources with a 24 micron counterpart, highlighting broad line AGN, and the Compton thick sources we find here. However, no clear separation of these sources can been seen.
 
\begin{figure}
\includegraphics[width=90mm]{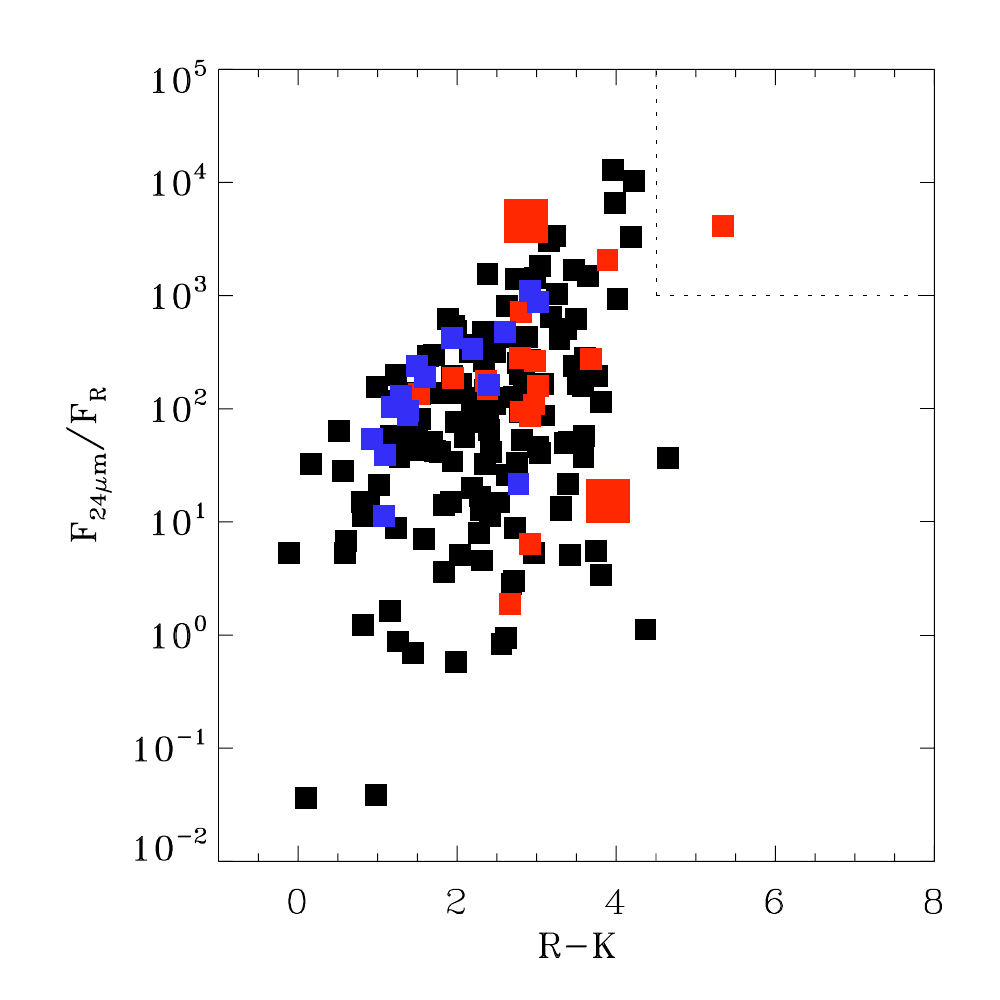}

\caption{The selection method of Fiore, et al (2008) for selecting Compton thick candidates given red R-K colours and a MIR excess (selection criteria shown as a dotted box). Blue squares are optical broad line AGN, and red squares are the Compton thick AGN we find here.  The largest red squares are Compton thick quasars with an intrinsic \lx$>10^{44}$ \ergs.}
 \label{fig_fiore}
\end{figure}

Secondly we explore the selection method of \cite{fiore08}, whereby red sources in R-K (R-K $>$ 4.5) with a MIR excess with respect to the optical ($F_{24\umu m}/F_{\rm{R}}>1000$) are identified as Compton thick candidates. We take R band WFI photometry from \cite{arnouts01} and K$_{\rm S}$ band SOFI photometry from \cite{olsen06}. Fig. \ref{fig_fiore} shows these selection criteria and where the sources we study here lie. This method has been applied to undetected  X-ray sources, and while only one source would be selected as being Compton thick using this method, it is indeed Compton thick as shown in our analysis.

\begin{figure}
\includegraphics[width=90mm]{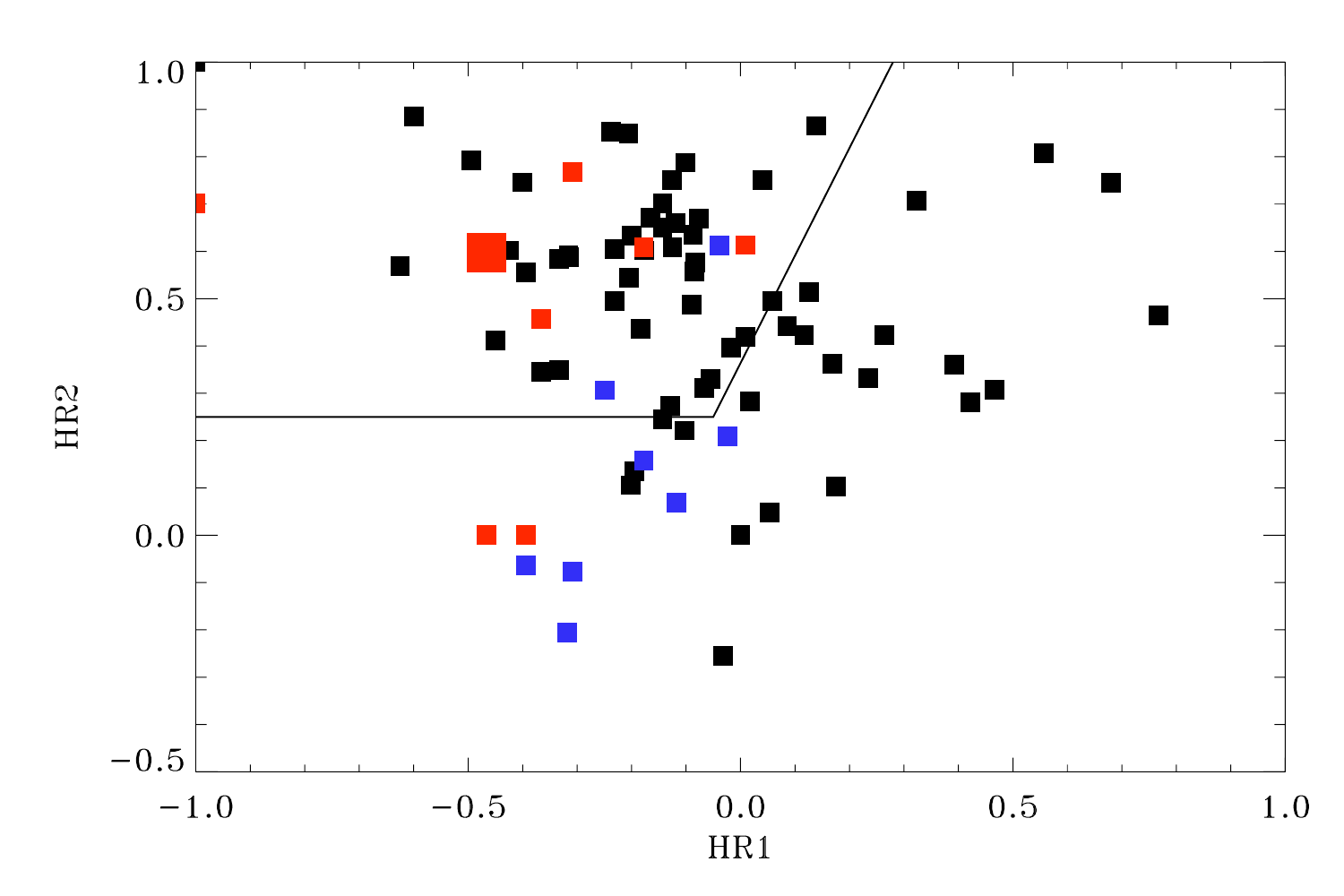}

\caption{The selection method of Brightman \& Nandra (2012), which uses HR1 (F(2-4 keV)-F(1-2 keV)/F(2-4 keV)+F(1-2 keV)) and HR2 (F(4-16 keV)-F(2-4 keV)/F(4-16 keV)+F(2-4 keV)) to select heavily obscured sources, where the box has been shown to include all Compton thick AGN. Shown for CDFS sources with $0.8<z<1.2$. Blue squares are optical broad line AGN, and red squares are the Compton thick AGN we find here.  The largest red squares are Compton thick quasars with an intrinsic \lx$>10^{44}$ \ergs.}
 \label{fig_hardness}
\end{figure}

Finally, the use of hardness ratios has often been used to infer the level of obscuration occurring when there are not enough counts for spectral fitting. These are computed using the number of counts detected in the hard band (H, typically 2-10 keV) and in the soft band (S, typically 0.5-2 keV), the hardness ratio then being H-S/H+S. This is of course very crude and does not account for spectral complexity, such as soft excess emission. \cite{brightman12} have shown that a pair of rest frame hardness ratios, HR1 (2-4/1-2 keV) and HR2 (4-16/2-4), do very well at picking out very obscured sources (\nh$>10^{23}$ \cmsq) which are hard in HR2 due to absorption, and soft in HR1 due to the soft excess emission, which we have found to be ubiquitous here in this analysis. Figure \ref{fig_hardness} shows this selection method for the CDFS sources with $0.8<z<1.2$, where all the bands used in this scheme are observed in the \chandra\ band. The solid wedge shows where all Compton thick sources are expected to lie. The separation of Compton thick sources and BLAGN here is much cleaner than with the multi-wavelength approaches.

\section{discussion}
\label{discussion}

\subsection{Geometrically buried AGN}

We find a significant number ($\sim20$\%) of heavily absorbed sources  in the CDFS in which evidence suggests that the central engine is buried in a close to 4$\pi$ steradians of material. These sources are best fit by a torus model with a 0\degree\ opening angle. Such geometrically buried AGN were predicted to be a large fraction of the local AGN population by \cite{ueda07} from the discovery of two such sources in a then, very small {\it Swift/BAT} sample. They reach this conclusion from the measurement of scattered nuclear light, which is believed to be scattered by gas which fills the cone of the torus, being a very small fraction of the intrinsic emission (\fscat\ $<0.5$\%). Due to the higher redshift nature of our sources, we cannot place such strict upper limits on the scattered fraction as the soft X-rays get redshifted out of the \chandra\ band. However we do find five sources in which we can place an upper limit of \fscat\ to be less than 0.5\%. 

Earlier work by \cite{levenson02} suggested the existence of AGN with tori that have small opening angles, using the equivalent width of the Fe K$\alpha$ as a diagnostic, but degeneracies in the parameters meant they could not confirm this assertion. Their sources were found to be in composite starburst/AGN sources, supporting the proposal by \cite{fabian98} which suggests that starbursts in the inner 100 pc of AGN are responsible for the nuclear obscuration in low luminosity AGN. Studies by \cite{noguchi09} and \cite{noguchi10} find that 25\% of obscured AGN in the \xmm\ Serendipitous Source Catalogue are geometrically buried. Further evidence for absorbed AGN with very low \fscat\ values comes from \cite{eguchi09} with {\it Suzaku} observations of six local AGN and \cite{comastri10} with {\it Suzaku} observations of five local AGN. \cite{comastri10} discuss the implications of these low \fscat\ values with respect to the XRB, concluding that the fits to the XRB spectrum  and source counts are essentially unaffected. Other works that have found evidence for buried AGN include \cite{rrobinson09} who suggest candidates for AGN with close to 100\% covering fractions in the Lockman Hole from sources with high torus dust luminosities with respect to their X-ray luminosities.

In this work, we attribute all soft excess emission in obscured sources to scattered nuclear X-rays. However, soft excess emission in obscured AGN can also be attributed to emission by circumnuclear gas photoionised by the AGN \citep{guainazzi07}, or to star-formation related processes in the host galaxy, such as X-ray binaries. Due to the spectral quality of our data we are unlikely to be able to discriminate between the different emission components. Nevertheless our conclusion from \fscat\ comes from its variation with \nh\ which is unlikely to be affected by these contaminants as photoionised emission is also likely to depend on the covering factor of the torus. While there have been suggestions that star formation and AGN obscuration are linked, the expected result from this is that \fscat\ would increase with \nh, rather than decrease.

Our data show that the more obscured along the line of sight an AGN is, the more likely it is to be buried in a geometrically thick torus, shown by small scattered fraction in these sources, and the preference to be fit by torus models with small opening angles. A similar conclusion was reached by \cite{ricci11}, who found from the study of local Seyfert galaxies using hard X-ray data from {\it INTEGRAL}, that Seyfert 2s with the strongest absorption also show the strongest reflection with respect to the less absorbed Seyferts, which they attribute to a larger covering fraction of the torus in the most obscured sources. Our data does not however show that geometrically buried AGN are more prevalent at lower luminosities, or that the opening angle of the torus is dependant on luminosity. This would be expected when considering the 'receding torus' model \citep{lawrence91, simpson05}, which predicts that as luminosity increases, the covering factor of the torus decreases as luminosity increases due to the radius of dust sublimation increasing and has much observational evidence \citep{ueda03, hasinger08, burlon11, brightman11b}. However, with the limited number of torus fits we have at hand, it may not be possible to break all degeneracies between \nh, \lx\ and redshift. 

As noted by \cite{elitzur12}, the variation in torus covering factors observed in AGN has important implications for the AGN unification scheme. This scheme employs only the orientation angle of the observer to explain the differences seen in AGN types \citep{antonucci93, urry95}, however it is now clear that the covering factor of the torus is also an important parameter. Open issues with AGN obscuration and unification are also reviewed in \cite{bianchi12}, including the discussion of the now apparent need for a model to explain the complex phenomenology observed in AGN. We have shown here that X-ray spectral analysis can give valuable information about the circumnuclear material in AGN, such as the covering factor, when the appropriate spectral models, such as those utilised in this paper, are exploited, and should be used in future to aid in a better understanding of AGN unification. 

\subsection{Compton thick AGN}

Of the 20 secure Compton thick sources in the CDFS we find in this investigation, source numbers 137, 178, 180, 186 and 282 were identified as Compton thick candidates by \cite{tozzi06} and 96, 137 and 186 were identified as Compton thick candidates by \cite{georgantopoulos07}, both with 1Ms \chandra\ data. Furthermore, source number 309 was presented as a Compton thick source by \cite{feruglio11} using the 4Ms data. Meanwhile, \cite{comastri11} presented two Compton thick sources from their analysis of ultra-deep \xmm\ data in the CDFS, which are source numbers 176 and 265 in the \cite{luo08} catalogue. From our spectral fits to the \chandra\ spectra we find that for 176 \nh=$6.57_{-1.21}^{+1.08}\times10^{23}$ \cmsq\ and 265 \nh=$1.04_{-0.17}^{+0.37}\times10^{24}$ \cmsq. Our result for 265 is in very good agreement   ($1.0_{-0.1}^{+0.2}\times10^{24}$ \cmsq) to the fit to the \xmm\ spectrum of the same source with the torus model of \cite{murphy09} by \cite{comastri11}. Of the remaining 21 1-$\sigma$ sources, 262, 265, 313 and 337 were identified as Compton thick candidates by \cite{tozzi06} and 262, 313 and 337  were identified as Compton thick candidates by \cite{georgantopoulos07}. Thus we are presenting 29 sources here as Compton thick which have to our knowledge, not been identified through X-ray spectroscopy previously.

We have accounted for the fraction of Compton thick sources which have not been identified through our spectral fitting, along with contaminating sources, with data from simulations. Furthermore, we take into account the selection bias against Compton thick sources due to extreme flux suppression in the X-ray band, which we do by restricting the intrinsic luminosity ranges considered to those which we know the survey is sensitive to Compton thick AGN. We find that the Compton thick fraction in the CDFS is a moderately increasing function of redshift. The fact that the Compton thick fraction does depend on redshift is consistent with the suggestion from \cite{treister09} that Compton thick sources have a steeper evolution from high redshift than less obscured sources. However, these authors report a much stronger increase in the Compton thick fraction of luminous AGN at $z>2$ based on X-ray stacking analysis of infrared excess sources. Our conclusion is also consistent with results that find the obscured fraction of AGN (\nh$>10^{22}$ \cmsq) increases with redshift \citep{lafranca05, ballantyne06, treister06, hasinger08}. It is also quantitatively consistent with the $(z+1)^{0.62}$ relation that \cite{hasinger08} finds for $f_{\rm 2}$ vs redshift, given the relation between $f_{\rm 2}$ and $f_{\rm CT}$ that we have assumed. The Compton thick fraction in the local universe ($z<0.1$) has been found to be $\approx$20\% \citep{brightman11b, burlon11} from both mid-infrared selected samples and hard X-ray {\it Swift/BAT} samples and our results show that this increases to $\approx$40\% at z=1-4. Our analyses include sources with \nh$>10^{25}$ \cmsq, which Chandra detects $<$8 keV through the scattered soft X-ray emission or reflected emission. Swift/BAT is less sensitive to these sources due to extreme flux suppression in the $>$10 keV band. However the results from the MIR selected sample of \cite{brightman11b}, which should be less biased against these extremely obscured sources, corroborate the hard X-ray detected results.

The normalised intrinsic Compton thick fractions that we have calculated are dependant on the $f_{\rm CT}$ vs. \lx\ relation that we have derived from the $f_{\rm 2}$ vs. \lx\ relation presented in \cite{hasinger08}. The assumption that we have made in deriving this relation is that  the number of sources with $10^{24}<$\nh$<10^{26}$ \cmsq\  is equal to that with $10^{21.5}<$\nh$<10^{24}$ \cmsq. If we were to change this and assume that the number of sources with $10^{24}<$\nh$<10^{26}$ \cmsq\  is only half that of $10^{21.5}<$\nh$<10^{24}$ \cmsq, then our normalised Compton thick fractions would change by less that 4\%, which is much less than the statistical error.

Several authors have presented results on work to constrain the space density of Compton thick AGN at high redshifts, particularly at z$\approx$2, many of which include stacking of X-ray undetected sources. Of these works, our results compare well with those of \cite{fiore08}, who find that the space density of infrared selected sources is in agreement with the XRB predictions of GCH07. Analyses by \cite{daddi07} have on the other hand over predicted the space density with respect to the XRB models. \cite{alexander11} have placed X-ray spectral constraints on X-ray detected z$\approx$2 $BzK$ selected galaxies in the CDFS and find that the Compton thick space density at these redshifts is lower than that predicted by the XRB.

We also briefly investigate alternative selection methods of Compton thick AGN, including multiwavelength selection. We find that Compton thick sources cannot be cleanly separated from the non Compton thick sources on a $F_{\rm X}-F_{\rm24 \umu m}$ diagram as they occupy the same regions of the diagram as BLAGN. \cite{georgantopoulos11} explore the use of the $L_{\rm X}-L_{\rm6 \umu m}$ ratio as an effective way of selecting Compton thick sources, and came to a similar conclusion, that this method cannot provide complete Compton thick AGN samples. While the selection method of \cite{fiore08} is aimed at X-ray undetected sources, it is clearly not suitable for sources detected in X-ray as only one source in the 2Ms catalogue of \cite{luo08} is selected as being Compton thick by their selection criteria. A scheme by \cite{brightman12} based on hardness ratios offers a promising alternative to spectral fitting for selecting heavily obscured sources.

\section{Conclusions}
We have applied new X-ray spectral models which account for Compton scattering and the geometry of the circumnuclear material from \cite{brightman11} to spectra of high redshift AGN in the \chandra\ Deep Field-South. Our conclusions from this investigation are as follows:
\begin{itemize}

\item We find evidence that the covering factor of the obscuring material and the line of sight \nh\ are linked, hence the most heavily obscured sources are also the most geometrically buried. We find this from the decline of the scattered fraction, \fscat\ with \nh, and from the prevalence of the most heavily obscured sources to be best fit by the torus model with the smallest opening angle.

\item We find that a significant number of heavily obscured (\nh$>10^{23}$ \cmsq) sources ($\sim20$\%) in the CDFS are buried in geometrically thick material, having been fit best with the torus model with a 0\degree\ opening angle. 

\item The average scattered fraction for sources in the CDFS is $1.7\pm0.4$\%.

\item We find a total of 41 Compton thick AGN in the CDFS from our analysis, 29 of which have not been identified as Compton thick by previous work on X-ray spectral analysis.

\item Binning our sample by redshift, considering only sources with an intrinsic luminosity greater than that to which the CDFS is sensitive, we find that the intrinsic Compton thick fraction, normalised to \lx$=10^{43.5}$ \ergs\ increases from $\approx20$\% in the local universe to $\approx40$\% at redshift 1-4.
 
\end{itemize} 

\section{Acknowledgements}

The authors would like to thank the anonymous referee for the careful reading and constructive criticism of our manuscript. MB would like to acknowledge the financial support from the Japan Society for the Promotion of Science (JSPS). Thanks also go to Elise Laird and Cyprian Rangel for providing reduced data in the CDFS, to Patrick Broos for help with the Acis Extract software, Bin Luo for providing solid angle data for the CDFS, and to John Silverman for useful discussions. We also thanks the builders and operators of \chandra. This research has made use of software provided by the Chandra X-ray Center (CXC) in the application package CIAO.

\bibliographystyle{mn2e}
\bibliography{bibdesk}

\begin{thebibliography}{}

\bibitem[\protect\citeauthoryear{{Alexander}, {Bauer}, {Brandt}, {Daddi},
  {Hickox}, {Lehmer}, {Luo}, {Xue}, {Young}, {Comastri}, {Del Moro}, {Fabian},
  {Gilli}, {Goulding}, {Mainieri}, {Mullaney}, {Paolillo} \&
  {Rafferty}}{{Alexander} et~al.}{2011}]{alexander11}
{Alexander} D.~M.,  {Bauer} F.~E.,  {Brandt} W.~N.,  {Daddi} E.,  {Hickox}
  R.~C.,  {Lehmer} B.~D.,  {Luo} B.,  {Xue} Y.~Q.,  {Young} M.,  {Comastri} A.,
   {Del Moro} A.,  {Fabian} A.~C.,  {Gilli} R.,  {Goulding} A.~D.,  {Mainieri}
  V.,  {Mullaney} J.~R.,  {Paolillo} M.,    {Rafferty} D.~A.,  2011, \apj, 738,
  44

\bibitem[\protect\citeauthoryear{{Antonucci}}{{Antonucci}}{1993}]{antonucci93}
{Antonucci} R.,  1993, ARA\&A, 31, 473

\bibitem[\protect\citeauthoryear{{Arnouts}, {Vandame}, {Benoist},
  {Groenewegen}, {da Costa}, {Schirmer}, {Mignani}, {Slijkhuis},
  {Hatziminaoglou}, {Hook}, {Madejsky}, {Rit{\'e}} \& {Wicenec}}{{Arnouts}
  et~al.}{2001}]{arnouts01}
{Arnouts} S.,  {Vandame} B.,  {Benoist} C.,  {Groenewegen} M.~A.~T.,  {da
  Costa} L.,  {Schirmer} M.,  {Mignani} R.~P.,  {Slijkhuis} R.,
  {Hatziminaoglou} E.,  {Hook} R.,  {Madejsky} R.,  {Rit{\'e}} C.,    {Wicenec}
  A.,  2001, \aap, 379, 740

\bibitem[\protect\citeauthoryear{{Awaki}, {Koyama}, {Inoue} \&
  {Halpern}}{{Awaki} et~al.}{1991}]{awaki91}
{Awaki} H.,  {Koyama} K.,  {Inoue} H.,    {Halpern} J.~P.,  1991, \pasj, 43,
  195

\bibitem[\protect\citeauthoryear{{Awaki}, {Terashima}, {Higaki} \&
  {Fukazawa}}{{Awaki} et~al.}{2009}]{awaki09}
{Awaki} H.,  {Terashima} Y.,  {Higaki} Y.,    {Fukazawa} Y.,  2009, \pasj, 61,
  317

\bibitem[\protect\citeauthoryear{{Balestra}, {Mainieri}, {Popesso},
  {Dickinson}, {Nonino}, {Rosati}, {Teimoorinia}, {Vanzella}, {Cristiani},
  {Cesarsky}, {Fosbury}, {Kuntschner} \& {Rettura}}{{Balestra}
  et~al.}{2010}]{balestra10}
{Balestra} I.,  {Mainieri} V.,  {Popesso} P.,  {Dickinson} M.,  {Nonino} M.,
  {Rosati} P.,  {Teimoorinia} H.,  {Vanzella} E.,  {Cristiani} S.,  {Cesarsky}
  C.,  {Fosbury} R.~A.~E.,  {Kuntschner} H.,    {Rettura} A.,  2010, \aap, 512,
  A12+

\bibitem[\protect\citeauthoryear{{Ballantyne}, {Everett} \&
  {Murray}}{{Ballantyne} et~al.}{2006}]{ballantyne06}
{Ballantyne} D.~R.,  {Everett} J.~E.,    {Murray} N.,  2006, \apj, 639, 740

\bibitem[\protect\citeauthoryear{{Bianchi}, {Maiolino} \& {Risaliti}}{{Bianchi}
  et~al.}{2012}]{bianchi12}
{Bianchi} S.,  {Maiolino} R.,    {Risaliti} G.,  2012, Advances in Astronomy,
  2012

\bibitem[\protect\citeauthoryear{{Brightman} \& {Nandra}}{{Brightman} \&
  {Nandra}}{2011a}]{brightman11}
{Brightman} M.,  {Nandra} K.,  2011a, \mnras, 413, 1206

\bibitem[\protect\citeauthoryear{{Brightman} \& {Nandra}}{{Brightman} \&
  {Nandra}}{2011b}]{brightman11b}
{Brightman} M.,  {Nandra} K.,  2011b, \mnras, 414, 3084

\bibitem[\protect\citeauthoryear{{Brightman} \& {Nandra}}{{Brightman} \&
  {Nandra}}{2012}]{brightman12}
{Brightman} M.,  {Nandra} K.,  2012, ArXiv e-prints

\bibitem[\protect\citeauthoryear{{Broos}, {Townsley}, {Feigelson}, {Getman},
  {Bauer} \& {Garmire}}{{Broos} et~al.}{2010}]{broos10}
{Broos} P.~S.,  {Townsley} L.~K.,  {Feigelson} E.~D.,  {Getman} K.~V.,  {Bauer}
  F.~E.,    {Garmire} G.~P.,  2010, \apj, 714, 1582

\bibitem[\protect\citeauthoryear{{Burlon}, {Ajello}, {Greiner}, {Comastri},
  {Merloni} \& {Gehrels}}{{Burlon} et~al.}{2011}]{burlon11}
{Burlon} D.,  {Ajello} M.,  {Greiner} J.,  {Comastri} A.,  {Merloni} A.,
  {Gehrels} N.,  2011, \apj, 728, 58

\bibitem[\protect\citeauthoryear{{Cash}}{{Cash}}{1979}]{cash79}
{Cash} W.,  1979, \apj, 228, 939

\bibitem[\protect\citeauthoryear{{Comastri}, {Iwasawa}, {Gilli}, {Vignali},
  {Ranalli}, {Matt} \& {Fiore}}{{Comastri} et~al.}{2010}]{comastri10}
{Comastri} A.,  {Iwasawa} K.,  {Gilli} R.,  {Vignali} C.,  {Ranalli} P.,
  {Matt} G.,    {Fiore} F.,  2010, \apj, 717, 787

\bibitem[\protect\citeauthoryear{{Comastri}, {Ranalli}, {Iwasawa}, {Vignali},
  {Gilli}, {Georgantopoulos}, {Barcons}, {Brandt}, {Brunner}, {Brusa},
  {Cappelluti}, {Carrera}, {Civano}, {Fiore}, {Hasinger}, {Mainieri} \&
  {Merloni}}{{Comastri} et~al.}{2011}]{comastri11}
{Comastri} A.,  {Ranalli} P.,  {Iwasawa} K.,  {Vignali} C.,  {Gilli} R.,
  {Georgantopoulos} I.,  {Barcons} X.,  {Brandt} W.~N.,  {Brunner} H.,  {Brusa}
  M.,  {Cappelluti} N.,  {Carrera} F.~J.,  {Civano} F.,  {Fiore} F.,
  {Hasinger} G.,  {Mainieri} V.,    {Merloni} A.,  2011, \aap, 526, L9+

\bibitem[\protect\citeauthoryear{{Comastri}, {Setti}, {Zamorani} \&
  {Hasinger}}{{Comastri} et~al.}{1995}]{comastri95}
{Comastri} A.,  {Setti} G.,  {Zamorani} G.,    {Hasinger} G.,  1995, \aap, 296,
  1

\bibitem[\protect\citeauthoryear{{Daddi}, {Alexander}, {Dickinson}, {Gilli},
  {Renzini}, {Elbaz}, {Cimatti}, {Chary}, {Frayer}, {Bauer}, {Brandt},
  {Giavalisco}, {Grogin}, {Huynh}, {Kurk}, {Mignoli}, {Morrison}, {Pope} \&
  {Ravindranath}}{{Daddi} et~al.}{2007}]{daddi07}
{Daddi} E.,  {Alexander} D.~M.,  {Dickinson} M.,  {Gilli} R.,  {Renzini} A.,
  {Elbaz} D.,  {Cimatti} A.,  {Chary} R.,  {Frayer} D.,  {Bauer} F.~E.,
  {Brandt} W.~N.,  {Giavalisco} M.,  {Grogin} N.~A.,  {Huynh} M.,  {Kurk} J.,
  {Mignoli} M.,  {Morrison} G.,  {Pope} A.,    {Ravindranath} S.,  2007, \apj,
  670, 173

\bibitem[\protect\citeauthoryear{{Donato}, {Sambruna} \& {Gliozzi}}{{Donato}
  et~al.}{2005}]{donato05}
{Donato} D.,  {Sambruna} R.~M.,    {Gliozzi} M.,  2005, \aap, 433, 1163

\bibitem[\protect\citeauthoryear{{Draper} \& {Ballantyne}}{{Draper} \&
  {Ballantyne}}{2011}]{draper11}
{Draper} A.~R.,  {Ballantyne} D.~R.,  2011, ArXiv e-prints

\bibitem[\protect\citeauthoryear{{Eguchi}, {Ueda}, {Awaki}, {Aird}, {Terashima}
  \& {Mushotzky}}{{Eguchi} et~al.}{2011}]{eguchi11}
{Eguchi} S.,  {Ueda} Y.,  {Awaki} H.,  {Aird} J.,  {Terashima} Y.,
  {Mushotzky} R.,  2011, \apj, 729, 31

\bibitem[\protect\citeauthoryear{{Eguchi}, {Ueda}, {Terashima}, {Mushotzky} \&
  {Tueller}}{{Eguchi} et~al.}{2009}]{eguchi09}
{Eguchi} S.,  {Ueda} Y.,  {Terashima} Y.,  {Mushotzky} R.,    {Tueller} J.,
  2009, \apj, 696, 1657

\bibitem[\protect\citeauthoryear{{Elitzur}}{{Elitzur}}{2012}]{elitzur12}
{Elitzur} M.,  2012, \apjl, 747, L33

\bibitem[\protect\citeauthoryear{{Fabian}, {Barcons}, {Almaini} \&
  {Iwasawa}}{{Fabian} et~al.}{1998}]{fabian98}
{Fabian} A.~C.,  {Barcons} X.,  {Almaini} O.,    {Iwasawa} K.,  1998, \mnras,
  297, L11+

\bibitem[\protect\citeauthoryear{{Fabian} \& {Iwasawa}}{{Fabian} \&
  {Iwasawa}}{1999}]{fabian99}
{Fabian} A.~C.,  {Iwasawa} K.,  1999, MNRAS, 303, L34

\bibitem[\protect\citeauthoryear{{Feruglio}, {Daddi}, {Fiore}, {Alexander},
  {Piconcelli} \& {Malacaria}}{{Feruglio} et~al.}{2011}]{feruglio11}
{Feruglio} C.,  {Daddi} E.,  {Fiore} F.,  {Alexander} D.~M.,  {Piconcelli} E.,
    {Malacaria} C.,  2011, ArXiv e-prints

\bibitem[\protect\citeauthoryear{{Fiore}, {Grazian}, {Santini}, {Puccetti},
  {Brusa}, {Feruglio}, {Fontana}, {Giallongo}, {Comastri}, {Gruppioni},
  {Pozzi}, {Zamorani} \& {Vignali}}{{Fiore} et~al.}{2008}]{fiore08}
{Fiore} F.,  {Grazian} A.,  {Santini} P.,  {Puccetti} S.,  {Brusa} M.,
  {Feruglio} C.,  {Fontana} A.,  {Giallongo} E.,  {Comastri} A.,  {Gruppioni}
  C.,  {Pozzi} F.,  {Zamorani} G.,    {Vignali} C.,  2008, \apj, 672, 94

\bibitem[\protect\citeauthoryear{{Georgantopoulos}, {Akylas}, {Georgakakis} \&
  {Rowan-Robinson}}{{Georgantopoulos} et~al.}{2009}]{georgantopoulos09}
{Georgantopoulos} I.,  {Akylas} A.,  {Georgakakis} A.,    {Rowan-Robinson} M.,
  2009, \aap, 507, 747

\bibitem[\protect\citeauthoryear{{Georgantopoulos}, {Georgakakis} \&
  {Akylas}}{{Georgantopoulos} et~al.}{2007}]{georgantopoulos07}
{Georgantopoulos} I.,  {Georgakakis} A.,    {Akylas} A.,  2007, \aap, 466, 823

\bibitem[\protect\citeauthoryear{{Georgantopoulos}, {Rovilos}, {Akylas},
  {Comastri}, {Ranalli}, {Vignali}, {Balestra}, {Gilli} \&
  {Cappelluti}}{{Georgantopoulos} et~al.}{2011}]{georgantopoulos11}
{Georgantopoulos} I.,  {Rovilos} E.,  {Akylas} A.,  {Comastri} A.,  {Ranalli}
  P.,  {Vignali} C.,  {Balestra} I.,  {Gilli} R.,    {Cappelluti} N.,  2011,
  \aap, 534, A23

\bibitem[\protect\citeauthoryear{{Ghisellini}, {Haardt} \& {Matt}}{{Ghisellini}
  et~al.}{1994}]{ghisellini94}
{Ghisellini} G.,  {Haardt} F.,    {Matt} G.,  1994, MNRAS, 267, 743

\bibitem[\protect\citeauthoryear{{Giacconi}, {Zirm}, {Wang}, {Rosati},
  {Nonino}, {Tozzi}, {Gilli}, {Mainieri}, {Hasinger}, {Kewley}, {Bergeron},
  {Borgani}, {Gilmozzi}, {Grogin}, {Koekemoer}, {Schreier}, {Zheng} \&
  {Norman}}{{Giacconi} et~al.}{2002}]{giacconi02}
{Giacconi} R.,  {Zirm} A.,  {Wang} J.,  {Rosati} P.,  {Nonino} M.,  {Tozzi} P.,
   {Gilli} R.,  {Mainieri} V.,  {Hasinger} G.,  {Kewley} L.,  {Bergeron} J.,
  {Borgani} S.,  {Gilmozzi} R.,  {Grogin} N.,  {Koekemoer} A.,  {Schreier} E.,
  {Zheng} W.,    {Norman} C.,  2002, \apjs, 139, 369

\bibitem[\protect\citeauthoryear{{Gilli}, {Comastri} \& {Hasinger}}{{Gilli}
  et~al.}{2007}]{gilli07}
{Gilli} R.,  {Comastri} A.,    {Hasinger} G.,  2007, A\&A, 463, 79

\bibitem[\protect\citeauthoryear{{Gilli}, {Su}, {Norman}, {Vignali},
  {Comastri}, {Tozzi}, {Rosati}, {Stiavelli}, {Brandt}, {Xue}, {Luo},
  {Castellano}, {Fontana}, {Fiore}, {Mainieri} \& {Ptak}}{{Gilli}
  et~al.}{2011}]{gilli11}
{Gilli} R.,  {Su} J.,  {Norman} C.,  {Vignali} C.,  {Comastri} A.,  {Tozzi} P.,
   {Rosati} P.,  {Stiavelli} M.,  {Brandt} W.~N.,  {Xue} Y.~Q.,  {Luo} B.,
  {Castellano} M.,  {Fontana} A.,  {Fiore} F.,  {Mainieri} V.,    {Ptak} A.,
  2011, \apjl, 730, L28+

\bibitem[\protect\citeauthoryear{{Guainazzi} \& {Bianchi}}{{Guainazzi} \&
  {Bianchi}}{2007}]{guainazzi07}
{Guainazzi} M.,  {Bianchi} S.,  2007, MNRAS, 374, 1290

\bibitem[\protect\citeauthoryear{{Hasinger}}{{Hasinger}}{2008}]{hasinger08}
{Hasinger} G.,  2008, \aap, 490, 905

\bibitem[\protect\citeauthoryear{{Ikeda}, {Awaki} \& {Terashima}}{{Ikeda}
  et~al.}{2009}]{ikeda09}
{Ikeda} S.,  {Awaki} H.,    {Terashima} Y.,  2009, \apj, 692, 608

\bibitem[\protect\citeauthoryear{{Komatsu}, {Dunkley}, {Nolta}, {Bennett},
  {Gold}, {Hinshaw}, {Jarosik}, {Larson}, {Limon}, {Page}, {Spergel},
  {Halpern}, {Hill}, {Kogut}, {Meyer}, {Tucker}, {Weiland}, {Wollack} \&
  {Wright}}{{Komatsu} et~al.}{2009}]{komatsu09}
{Komatsu} E.,  {Dunkley} J.,  {Nolta} M.~R.,  {Bennett} C.~L.,  {Gold} B.,
  {Hinshaw} G.,  {Jarosik} N.,  {Larson} D.,  {Limon} M.,  {Page} L.,
  {Spergel} D.~N.,  {Halpern} M.,  {Hill} R.~S.,  {Kogut} A.,  {Meyer} S.~S.,
  {Tucker} G.~S.,  {Weiland} J.~L.,  {Wollack} E.,    {Wright} E.~L.,  2009,
  \apjs, 180, 330

\bibitem[\protect\citeauthoryear{{La Franca}, {Fiore}, {Comastri}, {Perola},
  {Sacchi}, {Brusa}, {Cocchia}, {Feruglio}, {Matt}, {Vignali}, {Carangelo},
  {Ciliegi}, {Lamastra}, {Maiolino}, {Mignoli}, {Molendi} \& {Puccetti}}{{La
  Franca} et~al.}{2005}]{lafranca05}
{La Franca} F.,  {Fiore} F.,  {Comastri} A.,  {Perola} G.~C.,  {Sacchi} N.,
  {Brusa} M.,  {Cocchia} F.,  {Feruglio} C.,  {Matt} G.,  {Vignali} C.,
  {Carangelo} N.,  {Ciliegi} P.,  {Lamastra} A.,  {Maiolino} R.,  {Mignoli} M.,
   {Molendi} S.,    {Puccetti} S.,  2005, \apj, 635, 864

\bibitem[\protect\citeauthoryear{{Laird}, {Nandra}, {Georgakakis}, {Aird},
  {Barmby}, {Conselice}, {Coil}, {Davis}, {Faber}, {Fazio}, {Guhathakurta},
  {Koo}, {Sarajedini} \& {Willmer}}{{Laird} et~al.}{2009}]{laird09}
{Laird} E.~S.,  {Nandra} K.,  {Georgakakis} A.,  {Aird} J.~A.,  {Barmby} P.,
  {Conselice} C.~J.,  {Coil} A.~L.,  {Davis} M.,  {Faber} S.~M.,  {Fazio}
  G.~G.,  {Guhathakurta} P.,  {Koo} D.~C.,  {Sarajedini} V.,    {Willmer}
  C.~N.~A.,  2009, \apjs, 180, 102

\bibitem[\protect\citeauthoryear{{Lawrence}}{{Lawrence}}{1991}]{lawrence91}
{Lawrence} A.,  1991, \mnras, 252, 586

\bibitem[\protect\citeauthoryear{{Le F{\`e}vre}, {Vettolani}, {Paltani},
  {Tresse}, {Zamorani}, {Le Brun}, {Moreau}, {Bottini}, {Maccagni}, {Picat},
  {Scaramella}, {Scodeggio}, {Zanichelli}, {Adami}, {Arnouts} \&
  {Bardelli}}{{Le F{\`e}vre} et~al.}{2004}]{lefevre04}
{Le F{\`e}vre} O.,  {Vettolani} G.,  {Paltani} S.,  {Tresse} L.,  {Zamorani}
  G.,  {Le Brun} V.,  {Moreau} C.,  {Bottini} D.,  {Maccagni} D.,  {Picat}
  J.~P.,  {Scaramella} R.,  {Scodeggio} M.,  {Zanichelli} A.,  {Adami} C.,
  {Arnouts} S.,    {Bardelli} S.,  2004, \aap, 428, 1043

\bibitem[\protect\citeauthoryear{{Leahy} \& {Creighton}}{{Leahy} \&
  {Creighton}}{1993}]{leahy93}
{Leahy} D.~A.,  {Creighton} J.,  1993, MNRAS, 263, 314

\bibitem[\protect\citeauthoryear{{Levenson}, {Krolik}, {{\.Z}ycki}, {Heckman},
  {Weaver}, {Awaki} \& {Terashima}}{{Levenson} et~al.}{2002}]{levenson02}
{Levenson} N.~A.,  {Krolik} J.~H.,  {{\.Z}ycki} P.~T.,  {Heckman} T.~M.,
  {Weaver} K.~A.,  {Awaki} H.,    {Terashima} Y.,  2002, \apjl, 573, L81

\bibitem[\protect\citeauthoryear{{Luo}, {Bauer}, {Brandt}, {Alexander},
  {Lehmer}, {Schneider}, {Brusa}, {Comastri}, {Fabian}, {Finoguenov}, {Gilli},
  {Hasinger}, {Hornschemeier}, {Koekemoer}, {Mainieri}, {Paolillo} \&
  {Rosati}}{{Luo} et~al.}{2008}]{luo08}
{Luo} B.,  {Bauer} F.~E.,  {Brandt} W.~N.,  {Alexander} D.~M.,  {Lehmer} B.~D.,
   {Schneider} D.~P.,  {Brusa} M.,  {Comastri} A.,  {Fabian} A.~C.,
  {Finoguenov} A.,  {Gilli} R.,  {Hasinger} G.,  {Hornschemeier} A.~E.,
  {Koekemoer} A.,  {Mainieri} V.,  {Paolillo} M.,    {Rosati} P.,  2008, \apjs,
  179, 19

\bibitem[\protect\citeauthoryear{{Luo}, {Brandt}, {Xue}, {Brusa}, {Alexander},
  {Bauer}, {Comastri}, {Koekemoer}, {Lehmer}, {Mainieri}, {Rafferty},
  {Schneider}, {Silverman} \& {Vignali}}{{Luo} et~al.}{2010}]{luo10}
{Luo} B.,  {Brandt} W.~N.,  {Xue} Y.~Q.,  {Brusa} M.,  {Alexander} D.~M.,
  {Bauer} F.~E.,  {Comastri} A.,  {Koekemoer} A.,  {Lehmer} B.~D.,  {Mainieri}
  V.,  {Rafferty} D.~A.,  {Schneider} D.~P.,  {Silverman} J.~D.,    {Vignali}
  C.,  2010, \apjs, 187, 560

\bibitem[\protect\citeauthoryear{{Mainieri}, {Hasinger}, {Cappelluti}, {Brusa},
  {Brunner}, {Civano}, {Comastri}, {Elvis}, {Finoguenov}, {Fiore}, {Gilli},
  {Lehmann}, {Silverman}, {Tasca}, {Vignali}, {Zamorani}, {Schinnerer} \&
  {Impey}}{{Mainieri} et~al.}{2007}]{mainieri07}
{Mainieri} V.,  {Hasinger} G.,  {Cappelluti} N.,  {Brusa} M.,  {Brunner} H.,
  {Civano} F.,  {Comastri} A.,  {Elvis} M.,  {Finoguenov} A.,  {Fiore} F.,
  {Gilli} R.,  {Lehmann} I.,  {Silverman} J.,  {Tasca} L.,  {Vignali} C.,
  {Zamorani} G.,  {Schinnerer} E.,    {Impey} C.,  2007, \apjs, 172, 368

\bibitem[\protect\citeauthoryear{{Morrison} \& {McCammon}}{{Morrison} \&
  {McCammon}}{1983}]{morrison83}
{Morrison} R.,  {McCammon} D.,  1983, \apj, 270, 119

\bibitem[\protect\citeauthoryear{{Murphy} \& {Yaqoob}}{{Murphy} \&
  {Yaqoob}}{2009}]{murphy09}
{Murphy} K.~D.,  {Yaqoob} T.,  2009, \mnras, 397, 1549

\bibitem[\protect\citeauthoryear{{Nandra}, {O'Neill}, {George} \&
  {Reeves}}{{Nandra} et~al.}{2007}]{nandra07}
{Nandra} K.,  {O'Neill} P.~M.,  {George} I.~M.,    {Reeves} J.~N.,  2007,
  MNRAS, 382, 194

\bibitem[\protect\citeauthoryear{{Nandra} \& {Pounds}}{{Nandra} \&
  {Pounds}}{1994}]{nandra94}
{Nandra} K.,  {Pounds} K.~A.,  1994, MNRAS, 268, 405

\bibitem[\protect\citeauthoryear{{Noguchi}, {Terashima} \& {Awaki}}{{Noguchi}
  et~al.}{2009}]{noguchi09}
{Noguchi} K.,  {Terashima} Y.,    {Awaki} H.,  2009, \apj, 705, 454

\bibitem[\protect\citeauthoryear{{Noguchi}, {Terashima}, {Ishino}, {Hashimoto},
  {Koss}, {Ueda} \& {Awaki}}{{Noguchi} et~al.}{2010}]{noguchi10}
{Noguchi} K.,  {Terashima} Y.,  {Ishino} Y.,  {Hashimoto} Y.,  {Koss} M.,
  {Ueda} Y.,    {Awaki} H.,  2010, \apj, 711, 144

\bibitem[\protect\citeauthoryear{{Olsen}, {Miralles}, {da Costa}, {Madejsky},
  {J{\o}rgensen}, {Mignano}, {Arnouts}, {Benoist}, {Dietrich}, {Slijkhuis} \&
  {Zaggia}}{{Olsen} et~al.}{2006}]{olsen06}
{Olsen} L.~F.,  {Miralles} J.-M.,  {da Costa} L.,  {Madejsky} R.,
  {J{\o}rgensen} H.~E.,  {Mignano} A.,  {Arnouts} S.,  {Benoist} C.,
  {Dietrich} J.~P.,  {Slijkhuis} R.,    {Zaggia} S.,  2006, \aap, 456, 881

\bibitem[\protect\citeauthoryear{{Page}, {Carrera}, {Stevens}, {Ebrero} \&
  {Blustin}}{{Page} et~al.}{2011}]{page11}
{Page} M.~J.,  {Carrera} F.~J.,  {Stevens} J.~A.,  {Ebrero} J.,    {Blustin}
  A.~J.,  2011, \mnras, pp 1235--+

\bibitem[\protect\citeauthoryear{{Popesso}, {Dickinson}, {Nonino}, {Vanzella},
  {Daddi}, {Fosbury}, {Kuntschner}, {Mainieri}, {Cristiani}, {Cesarsky},
  {Giavalisco}, {Renzini} \& {GOODS Team}}{{Popesso} et~al.}{2009}]{popesso09}
{Popesso} P.,  {Dickinson} M.,  {Nonino} M.,  {Vanzella} E.,  {Daddi} E.,
  {Fosbury} R.~A.~E.,  {Kuntschner} H.,  {Mainieri} V.,  {Cristiani} S.,
  {Cesarsky} C.,  {Giavalisco} M.,  {Renzini} A.,    {GOODS Team} 2009, \aap,
  494, 443

\bibitem[\protect\citeauthoryear{{Ravikumar}, {Puech}, {Flores}, {Proust},
  {Hammer}, {Lehnert}, {Rawat}, {Amram}, {Balkowski}, {Burgarella}, {Cassata},
  {Cesarsky}, {Cimatti}, {Combes}, {Daddi}, {Dannerbauer} \& {di Serego
  Alighieri}}{{Ravikumar} et~al.}{2007}]{ravikumar07}
{Ravikumar} C.~D.,  {Puech} M.,  {Flores} H.,  {Proust} D.,  {Hammer} F.,
  {Lehnert} M.,  {Rawat} A.,  {Amram} P.,  {Balkowski} C.,  {Burgarella} D.,
  {Cassata} P.,  {Cesarsky} C.,  {Cimatti} A.,  {Combes} F.,  {Daddi} E.,
  {Dannerbauer} H.,    {di Serego Alighieri} S.,  2007, \aap, 465, 1099

\bibitem[\protect\citeauthoryear{{Ricci}, {Walter}, {Courvoisier} \&
  {Paltani}}{{Ricci} et~al.}{2011}]{ricci11}
{Ricci} C.,  {Walter} R.,  {Courvoisier} T.~J.-L.,    {Paltani} S.,  2011,
  \aap, 532, A102

\bibitem[\protect\citeauthoryear{{Risaliti}, {Maiolino} \&
  {Salvati}}{{Risaliti} et~al.}{1999}]{risaliti99}
{Risaliti} G.,  {Maiolino} R.,    {Salvati} M.,  1999, ApJ, 522, 157

\bibitem[\protect\citeauthoryear{{Risaliti}, {Salvati}, {Elvis}, {Fabbiano},
  {Baldi}, {Bianchi}, {Braito}, {Guainazzi}, {Matt}, {Miniutti}, {Reeves},
  {Soria} \& {Zezas}}{{Risaliti} et~al.}{2009}]{risaliti09}
{Risaliti} G.,  {Salvati} M.,  {Elvis} M.,  {Fabbiano} G.,  {Baldi} A.,
  {Bianchi} S.,  {Braito} V.,  {Guainazzi} M.,  {Matt} G.,  {Miniutti} G.,
  {Reeves} J.,  {Soria} R.,    {Zezas} A.,  2009, \mnras, 393, L1

\bibitem[\protect\citeauthoryear{{Rivers}, {Markowitz} \&
  {Rothschild}}{{Rivers} et~al.}{2011}]{rivers11}
{Rivers} E.,  {Markowitz} A.,    {Rothschild} R.,  2011, \apj, 732, 36

\bibitem[\protect\citeauthoryear{{Rowan-Robinson}, {Valtchanov} \&
  {Nandra}}{{Rowan-Robinson} et~al.}{2009}]{rrobinson09}
{Rowan-Robinson} M.,  {Valtchanov} I.,    {Nandra} K.,  2009, \mnras, 397, 1326

\bibitem[\protect\citeauthoryear{{Silverman}, {Mainieri}, {Salvato},
  {Hasinger}, {Bergeron}, {Capak}, {Szokoly}, {Finoguenov}, {Gilli}, {Rosati},
  {Tozzi}, {Vignali}, {Alexander}, {Brandt}, {Lehmer}, {Luo}, {Rafferty} \&
  {Xue}}{{Silverman} et~al.}{2010}]{silverman10}
{Silverman} J.~D.,  {Mainieri} V.,  {Salvato} M.,  {Hasinger} G.,  {Bergeron}
  J.,  {Capak} P.,  {Szokoly} G.,  {Finoguenov} A.,  {Gilli} R.,  {Rosati} P.,
  {Tozzi} P.,  {Vignali} C.,  {Alexander} D.~M.,  {Brandt} W.~N.,  {Lehmer}
  B.~D.,  {Luo} B.,  {Rafferty} D.,    {Xue} Y.~Q.,  2010, \apjs, 191, 124

\bibitem[\protect\citeauthoryear{{Simpson}}{{Simpson}}{2005}]{simpson05}
{Simpson} C.,  2005, \mnras, 360, 565

\bibitem[\protect\citeauthoryear{{Szokoly}, {Bergeron}, {Hasinger}, {Lehmann},
  {Kewley}, {Mainieri}, {Nonino}, {Rosati}, {Giacconi}, {Gilli}, {Gilmozzi},
  {Norman}, {Romaniello}, {Schreier}, {Tozzi}, {Wang}, {Zheng} \&
  {Zirm}}{{Szokoly} et~al.}{2004}]{szokoly04}
{Szokoly} G.~P.,  {Bergeron} J.,  {Hasinger} G.,  {Lehmann} I.,  {Kewley} L.,
  {Mainieri} V.,  {Nonino} M.,  {Rosati} P.,  {Giacconi} R.,  {Gilli} R.,
  {Gilmozzi} R.,  {Norman} C.,  {Romaniello} M.,  {Schreier} E.,  {Tozzi} P.,
  {Wang} J.~X.,  {Zheng} W.,    {Zirm} A.,  2004, \apjs, 155, 271

\bibitem[\protect\citeauthoryear{{Tazaki}, {Ueda}, {Terashima} \&
  {Mushotzky}}{{Tazaki} et~al.}{2011}]{tazaki11}
{Tazaki} F.,  {Ueda} Y.,  {Terashima} Y.,    {Mushotzky} R.~F.,  2011, ArXiv
  e-prints

\bibitem[\protect\citeauthoryear{{Tozzi}, {Gilli}, {Mainieri}, {Norman},
  {Risaliti}, {Rosati}, {Bergeron}, {Borgani}, {Giacconi}, {Hasinger},
  {Nonino}, {Streblyanska}, {Szokoly}, {Wang} \& {Zheng}}{{Tozzi}
  et~al.}{2006}]{tozzi06}
{Tozzi} P.,  {Gilli} R.,  {Mainieri} V.,  {Norman} C.,  {Risaliti} G.,
  {Rosati} P.,  {Bergeron} J.,  {Borgani} S.,  {Giacconi} R.,  {Hasinger} G.,
  {Nonino} M.,  {Streblyanska} A.,  {Szokoly} G.,  {Wang} J.~X.,    {Zheng} W.,
   2006, A\&A, 451, 457

\bibitem[\protect\citeauthoryear{{Tran}}{{Tran}}{2003}]{tran03}
{Tran} H.~D.,  2003, ApJ, 583, 632

\bibitem[\protect\citeauthoryear{{Treister} \& {Urry}}{{Treister} \&
  {Urry}}{2006}]{treister06}
{Treister} E.,  {Urry} C.~M.,  2006, \apjl, 652, L79

\bibitem[\protect\citeauthoryear{{Treister}, {Urry} \& {Virani}}{{Treister}
  et~al.}{2009}]{treister09}
{Treister} E.,  {Urry} C.~M.,    {Virani} S.,  2009, \apj, 696, 110

\bibitem[\protect\citeauthoryear{{Turner}, {George}, {Nandra} \&
  {Mushotzky}}{{Turner} et~al.}{1997}]{turner97}
{Turner} T.~J.,  {George} I.~M.,  {Nandra} K.,    {Mushotzky} R.~F.,  1997,
  ApJS, 113, 23

\bibitem[\protect\citeauthoryear{{Ueda}, {Akiyama}, {Ohta} \& {Miyaji}}{{Ueda}
  et~al.}{2003}]{ueda03}
{Ueda} Y.,  {Akiyama} M.,  {Ohta} K.,    {Miyaji} T.,  2003, ApJ, 598, 886

\bibitem[\protect\citeauthoryear{{Ueda}, {Eguchi}, {Terashima}, {Mushotzky},
  {Tueller}, {Markwardt}, {Gehrels}, {Hashimoto} \& {Potter}}{{Ueda}
  et~al.}{2007}]{ueda07}
{Ueda} Y.,  {Eguchi} S.,  {Terashima} Y.,  {Mushotzky} R.,  {Tueller} J.,
  {Markwardt} C.,  {Gehrels} N.,  {Hashimoto} Y.,    {Potter} S.,  2007, ApJL,
  664, L79

\bibitem[\protect\citeauthoryear{{Urry} \& {Padovani}}{{Urry} \&
  {Padovani}}{1995}]{urry95}
{Urry} C.~M.,  {Padovani} P.,  1995, \pasp, 107, 803

\bibitem[\protect\citeauthoryear{{Wolf}, {Hildebrandt}, {Taylor} \&
  {Meisenheimer}}{{Wolf} et~al.}{2008}]{wolf08}
{Wolf} C.,  {Hildebrandt} H.,  {Taylor} E.~N.,    {Meisenheimer} K.,  2008,
  \aap, 492, 933

\bibitem[\protect\citeauthoryear{{Wolf}, {Meisenheimer}, {Kleinheinrich},
  {Borch}, {Dye}, {Gray}, {Wisotzki}, {Bell}, {Rix}, {Cimatti}, {Hasinger} \&
  {Szokoly}}{{Wolf} et~al.}{2004}]{wolf04}
{Wolf} C.,  {Meisenheimer} K.,  {Kleinheinrich} M.,  {Borch} A.,  {Dye} S.,
  {Gray} M.,  {Wisotzki} L.,  {Bell} E.~F.,  {Rix} H.-W.,  {Cimatti} A.,
  {Hasinger} G.,    {Szokoly} G.,  2004, \aap, 421, 913

\bibitem[\protect\citeauthoryear{{Wuyts}, {Labb{\'e}}, {Schreiber}, {Franx},
  {Rudnick}, {Brammer} \& {van Dokkum}}{{Wuyts} et~al.}{2008}]{wuyts08}
{Wuyts} S.,  {Labb{\'e}} I.,  {Schreiber} N.~M.~F.,  {Franx} M.,  {Rudnick} G.,
   {Brammer} G.~B.,    {van Dokkum} P.~G.,  2008, \apj, 682, 985

\bibitem[\protect\citeauthoryear{{Xue}, {Luo}, {Brandt}, {Bauer}, {Lehmer},
  {Broos}, {Schneider}, {Alexander}, {Brusa}, {Comastri}, {Fabian}, {Gilli},
  {Hasinger}, {Hornschemeier}, {Koekemoer}, {Liu} \& {Mainieri}}{{Xue}
  et~al.}{2011}]{xue11}
{Xue} Y.~Q.,  {Luo} B.,  {Brandt} W.~N.,  {Bauer} F.~E.,  {Lehmer} B.~D.,
  {Broos} P.~S.,  {Schneider} D.~P.,  {Alexander} D.~M.,  {Brusa} M.,
  {Comastri} A.,  {Fabian} A.~C.,  {Gilli} R.,  {Hasinger} G.,  {Hornschemeier}
  A.~E.,  {Koekemoer} A.,  {Liu} T.,    {Mainieri} V.,  2011, \apjs, 195, 10

\bibitem[\protect\citeauthoryear{{Yaqoob}}{{Yaqoob}}{1997}]{yaqoob97}
{Yaqoob} T.,  1997, ApJ, 479, 184

\bibitem[\protect\citeauthoryear{{Zheng}, {Mikles}, {Mainieri}, {Hasinger},
  {Rosati}, {Wolf}, {Norman}, {Szokoly}, {Gilli}, {Tozzi}, {Wang}, {Zirm} \&
  {Giacconi}}{{Zheng} et~al.}{2004}]{zheng04}
{Zheng} W.,  {Mikles} V.~J.,  {Mainieri} V.,  {Hasinger} G.,  {Rosati} P.,
  {Wolf} C.,  {Norman} C.,  {Szokoly} G.,  {Gilli} R.,  {Tozzi} P.,  {Wang}
  J.~X.,  {Zirm} A.,    {Giacconi} R.,  2004, \apjs, 155, 73

\end{thebibliography}

\label{lastpage}
\end{document}